\documentclass[11pt,a4paper]{article}
\usepackage{jheppub}
\usepackage{amsmath}
\usepackage{amssymb}
\usepackage{braket}
\usepackage{bbm}
\usepackage{verbatim}
\usepackage[normalem]{ulem}
\usepackage{bm}
\usepackage{natbib}
\usepackage{soul}
\usepackage{ulem}
\usepackage{cool}

\newcommand{\bU}{\bar{U}}

\newcommand{\be}{\begin{equation}}
\newcommand{\ee}{\end{equation}}
\newcommand{\ev}[1]{\langle #1 \rangle}

\title{Energy is Entanglement}

\author[a,b]{Stefan Leichenauer,}
\author[a,b]{Adam Levine,}
\author[a,b]{and Arvin Shahbazi-Moghaddam}

\affiliation[a]{Center for Theoretical Physics and Department of Physics,\\
University of California, Berkeley, CA 94720, U.S.A.} 
\affiliation[b]{Lawrence Berkeley National Laboratory, Berkeley, CA 94720, U.S.A.}

\abstract{We compute the local second variation of the von Neumann entropy of a region in theories with a gravity dual. For null variations our formula says that the diagonal part of the Quantum Null Energy Condition is saturated in every state, thus providing an equivalence between energy and entropy. We prove that the formula holds at leading order in $1/N$, and further argue that it will not be affected at higher orders. We conjecture that the QNEC is saturated in all interacting theories. We also discuss the special case of free theories, and the implications of our formula for the Averaged Null Energy Condition, Quantum Focusing Conjecture, and gravitational equations of motion. We show that the leading-order gravitational equations of motion, Einstein's equations, are equivalent to leading-order saturation of the QFC for Planck-width deformations.}

\begin{document}
\maketitle

\section{Introduction}\label{sec:intro}

The connection between quantum information and energy has been an emerging theme of recent progress in quantum field theory. Causality combined with universal inequalities like positivity and monotonicity of relative entropy can be used to derive many interesting energy-entropy bounds.  Examples include the Bekenstein bound~\cite{Casini:2008aa}, the quantum Bousso bound~\cite{Bousso:2014sda, Bousso:2015aa}, the Averaged Null Energy Condition (ANEC)~\cite{Faulkner:2016mzt, Hartman:2016lgu}, and the Quantum Null Energy Condition (QNEC)~\cite{Bousso:2016aa,Koeller:2015qmn, Balakrishnan:2017aa,Wall:2017aa}. Here we strengthen the energy-entropy connection, moving from bounds to equalities.

The key insight of the QNEC, which we will exploit, is that one should look at variations of the entropy $S$ of a region as the region is deformed. Consider the entropy as a functional of the entangling surface embedding functions $X^\mu$. Then one can compute the functional derivative $\delta^2 S/ \delta X^\mu(y) \delta X^\nu(y')$ which encodes how the entropy depends on the shape of the region. In general, this second variation will contain contact, or ``diagonal," terms, proportional to $\delta$-functions and derivatives of $\delta$-functions, as well as ``off-diagonal" terms. Our interest here is in the $\delta$-function contact term, and we introduce $S''_{\mu\nu}$ as the coefficient of the $\delta$-function:
\be\label{eq-variation1}
\frac{\delta^2 S}{\delta X^\mu(y) \delta X^\nu(y')} = S''_{\mu\nu}(y)\delta^{(d-2)}(y-y') + \cdots
\ee

\paragraph{Null Variations}

First consider the null-null component of the second variation, $S''_{vv}(y)$, where $v$ is a null coordinate in a direction orthogonal to the entangling surface at the point $y$.\footnote{We are restricting attention to field theories in Minkowski space throughout the main text.} Suppose the entangling surface is locally restricted to lie in the null plane orthogonal to $v$ near the point $y$. With this setup we can apply the QNEC, which says $S''_{vv} \leq 2\pi \ev{T_{vv}}$. Our main conjecture is that this inequality is always saturated:\footnote{In \cite{Ecker:2017aa} the issue of QNEC saturation was also investigated, but this is a different notion of saturation. Their analysis did not isolate the $\delta$-function component, and instead considered the total variation in the entropy including the contribution of off-diagonal terms. So the examples in \cite{Ecker:2017aa} where the QNEC is not ``saturated" are not in contradiction with our results.}
\be\label{eq-nullconjecture}
S''_{vv} = 2\pi \ev{T_{vv}}.
\ee
We believe this holds for all relativistic quantum field theories with an interacting UV fixed point in $d>2$ dimensions. For the special case of an interacting CFT this fully specifies the stress tensor in terms of entropy variations: by considering \eqref{eq-nullconjecture} for all entangling surfaces passing through a point, $\ev{T_{\mu\nu}}$ is completely determined up to a trace term. In a CFT the trace of the stress tensor vanishes, and so the entropy variations determine the full stress tensor in that case. This is the sense in which energy comes from entanglement.

Our primary evidence for \eqref{eq-nullconjecture} is holographic, as explained below. But if we restrict attention to quantities that can be built out of local expectation values of operators and the local surface geometry there is no other possibility for $S_{vv}''$. A significant constraint comes from considering the vacuum modular Hamiltonian, $K$, which is defined by
\be 
S(\sigma + \delta \sigma) - S(\sigma) = {\rm Tr}\left(K\delta \sigma\right) + O\!\left(\delta\sigma^2\right),
\ee
where $\sigma$ is the vacuum state reduced to the region under consideration and $\delta \sigma$ is an arbitrary perturbation of the state. If we had a general formula for $S$ in terms of expectation values of operators, we would be able to read off the modular Hamiltonian from the terms in that formula linear in expectation values.\footnote{For simplicity of the discussion we set all vacuum expectation values to zero.} For a region bounded by an entangling surface restricted to a null plane the modular Hamiltonian has a known formula in terms of the stress tensor~\cite{Casini:2017aa}, and in particular we have
\be\label{eq-modhamvar}
K''_{vv} = 2\pi \ev{T_{vv}}.
\ee
That is why $\ev{T_{vv}}$ is the only possible linear term we could have had in \eqref{eq-nullconjecture}.

A nonlinear contribution to $S_{vv}''$, such as a product of expectation values, is restricted by dimensional analysis and unitarity bounds: the only possibility is if the theory contains a free field. Then we can take the classical expression for $T_{vv}$, which is quadratic in the field, and replace each of those fields with expectation values to get an expression quadratic in expectation values with the right dimensionality to contribute to $S''_{vv}$. For interacting fields, nonzero anomalous dimensions prevent this from working. We will say more about free theories in Appendix~\ref{sec-free}, where we will see that this possibility is realized by a term $\sim \ev{\partial_v\phi}^2$ for a free scalar field, which is why we limit ourselves to interacting theories in the main text. The substance of \eqref{eq-nullconjecture}, then, is the statement that there are no non-local contributions to $S_{vv}''$.

\paragraph{Relative Entropy}
There is a natural interpretation of \eqref{eq-nullconjecture} in terms of relative entropy. The relative entropy of a state $\rho$ and a reference state $\sigma$---for us, the vacuum---is a measure of the distinguishability of the two states. We will denote the relative entropy of $\rho$ and the vacuum by $S_{\rm rel}(\rho)$. By definition, the relative entropy is 
\be
S_{\rm rel}(\rho)  = \Delta \ev{K} -\Delta S, 
\ee
where $\Delta \ev{K}$ and $\Delta S$ denote the vacuum-subtracted modular energy and vacuum-subtracted entropy, respectively.  A consequence of \eqref{eq-nullconjecture} is that $\Delta S_{vv}'' = \Delta \ev{K_{vv}''}$, so we can say that
\be
S_{{\rm rel}, vv}'' = 0.
\ee
This equation is implied by \eqref{eq-nullconjecture} but is weaker, since it does not require us to know what the modular Hamiltonian actually is. The extra information of \eqref{eq-nullconjecture} is the expression \eqref{eq-modhamvar} for the second variation of the modular Hamiltonian. It can be useful to formulate our results in terms of relative entropy instead of entropy itself because relative entropy is generally free from UV divergences, at least for nice states.\footnote{It is possible for relative entropy to be infinite, for instance if we take our region to be the whole space and consider two orthogonal pure states. This is an expected and understood type of infinity, and not dependent on a choice of UV regulator.}

\paragraph{Non-Null Deformations}
Now let us move beyond the null case. Our goal in doing this is to understand the simplest setup where non-null deformations can be analyzed, and so we will make several additional restrictions that we do not make in the null case. As explained in~\cite{Akers:2017aa, Fu:2017aa} and below in Section~\ref{sec-bulksetup}, \eqref{eq-nullconjecture} for the null case is a well-defined, finite equation in field theory. Local stationarity conditions on the entangling surface are enough to eliminate state-independent geometric divergences in the entropy, and the remaining state-dependent divergences cancel between the entropy and stress tensor. In the non-null case, eliminating divergences is more difficult. State-independent divergences can be dealt with by considering the vacuum-subtracted entropy $\Delta S$ rather than just $S$. State-dependent divergences associated with low-lying operators in the theory are more problematic. To eliminate these divergences, it is enough to restrict our attention theories where all relevant couplings have mass dimension greater than $d/2$, and to states where operators of dimension $\Delta \leq d/2$ have vanishing expectation values near the entangling surface. The idea of these restrictions is to make sure there are no parameters with scaling dimension small enough to contribute to divergences. We will make the further restriction in the non-null case to planar entangling surfaces, and this last restriction is made purely to simplify the analysis and presentation. With these assumptions in place we find
\be\label{eq-conjecture}
\Delta S''_{\mu\nu} = 2\pi \left( n_{\mu}^\rho n_\nu^\sigma \ev{T_{\rho \sigma}} + \frac{d^2-3d-2}{2(d+1)(d-2)}n_{\mu\nu} h^{ab} \ev{T_{ab}}\right),
\ee
where $n_{\mu\nu}$ is the normal projector to the entangling surface and $h_{ab}$ is the intrinsic metric on the entangling surface.\footnote{In \cite{Wall:2017aa}, a quantum version of the dominant energy condition which involved spacelike deformations of entropy was proposed for $d=2$ dimensions. In that inequality, timelike components of the stress tensor were bounded by spacelike components of the entropy variation, whereas in \eqref{eq-conjecture} timelike components of the stress tensor are related to timelike components of $\Delta S''_{\mu\nu}$ (ignoring the second term of \eqref{eq-conjecture}, which is absent in two dimensions). Our techniques are not directly applicable to two dimensions, and a na\"ive extrapolation of \eqref{eq-conjecture} is probably incorrect, but it would interesting to investigate this issue further in the future.} Note that \eqref{eq-conjecture} implies that $S_{{\rm rel},\mu\nu}'' = 0$.

\paragraph{Consequences for Field Theory and Gravity}

We view \eqref{eq-nullconjecture} and \eqref{eq-conjecture} as deep truths about interacting quantum field theories, worthy of further study. At present, our evidence for these conjectures comes from holography. We will calculate $S_{\mu\nu}''$ directly and prove that \eqref{eq-nullconjecture} and \eqref{eq-conjecture} hold precisely at leading order in large-$N$ for all bulk states. We will also argue that subleading corrections in $1/N$ do not alter these conclusions. While this does not amount to a full proof, it is enough evidence for us to posit that \eqref{eq-nullconjecture} is true universally, and that \eqref{eq-conjecture} holds with relatively few additional assumptions.

An immediate application, which we discuss in Section~\ref{sec:discussion}, is to gravity. If we couple our field theory to gravity, then we can effectively isolate the $\delta$-function part of the null second variation by deforming the entangling surface over a Planck-sized, or slightly larger, domain. According to the Raychaudhuri equation, if the surface is locally stationary then the leading change in its area due to this deformation is determined by $R_{vv}$, the null-null component of the Ricci tensor. Using \eqref{eq-nullconjecture} together with Einstein's equations, $R_{vv} = 8\pi G_N T_{vv}$, we learn that this change in area is precisely canceled by $4G_NS_{vv}''$. This means that the leading-order change in generalized entropy---area in Planck units plus entropy---is actually zero under such a deformation. In Section~\ref{sec:discussion} we will show how this argument can also be reversed, demonstrating that this leading-order cancellation in the variation of the generalized entropy can be taken as a fundamental principle and used to derive Einstein's equations. This is essentially an update of the thermodynamic derivation of Einstein's equations by Jacobson~\cite{Jacobson:1995aa}.

\paragraph{Outline} In Section~\ref{sec:2} we review some of the basic concepts of entropy, relative entropy, and the holographic setup that will be relevant for our calculation. In Section~\ref{sec:bulkpert} we prove \eqref{eq-nullconjecture} for situations where it is sufficient to consider linear perturbations of the bulk geometry. This includes any state where gravitational backreaction in the bulk is small. In Section~\ref{sec:genproof} we extend this proof to any bulk state. The idea is that $S''_{vv}$ is related to near-boundary physics in the bulk, and for any state the near-boundary geometry is approximately vacuum. So the proof reduces to the linear case. In Section~\ref{sec:nonnull} we move away from null deformations to prove \eqref{eq-conjecture} using the same techniques. We conclude in Section~\ref{sec:discussion} with a discussion of extensions and implications of our work. Several appendices are included discussing closely related topics.


\section{Setup and Conventions}\label{sec:2}

In this section we will make some general remarks about the known relations between entropy and energy, and the implications of our conjecture.


\subsection{The Field Theory Setup}\label{sec-ftsetup}

\begin{figure}
	\centering
	\includegraphics[scale=.7]{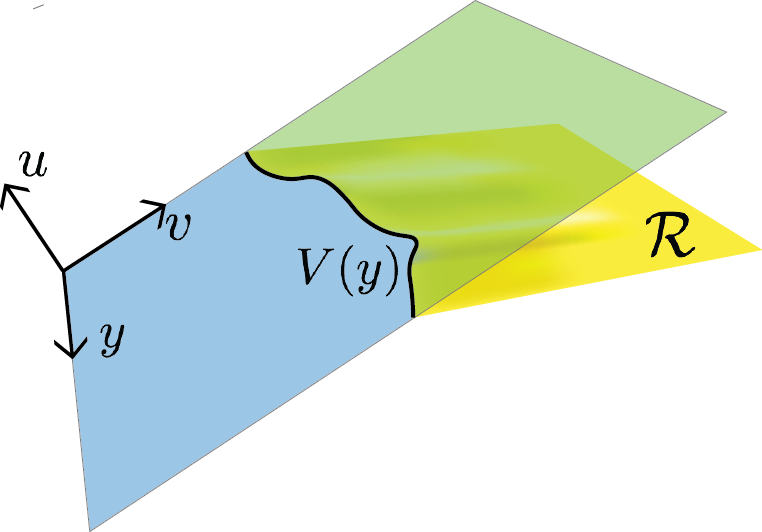}
	\caption{Most of our work concerns the variations of entanglement entropy for the yellow region $\mathcal{R}$ whose boundary $\partial \mathcal{R}$ lies on the null plane $u=0$. The entangling surface is specified by the function $V(y)$.}\label{fig:2}
\end{figure}

Let $u = (t-x)/\sqrt{2}$ and $v=(t+x)/\sqrt{2}$ be null coordinates, and let $y$ denote the other $d-2$ spatial coordinates. For now, and for most of the rest of the paper, we will take the boundary of our region $\partial \mathcal{R}$ to be a section of the null plane $u=0$. This boundary is specified by the equation $v=V(y)$. We take the region $\mathcal{R}$ to be a surface lying witin the ``right quadrant," having $u<0$ and $v>V(y)$ (marked in yellow in Fig \ref{fig:2}). A one-parameter family of functions $V_\lambda(y)$ specifies a one-parameter family of regions $\mathcal{R}(\lambda)$. We always take the one-parameter family to be of the form $V_\lambda(y) = V_0(y) + \lambda \dot V(y)$ with $\dot V \geq 0$, so that $\lambda$ plays the roll of an affine parameter along a future-directed null geodesic located at position $y$.

Given any global state of the theory, we can compute the von Neumann entropy $S$ of the region $\mathcal{R}$. Keeping the state fixed, the entropy becomes a functional of the boundary of the region, $S = S[V(y)]$. When we have a one-parameter family of regions, then we can write $S(\lambda) = S[V_\lambda(y)]$. Throughout the rest of this work we will be interested in the derivatives of $S$ with respect to $\lambda$, as well as the functional derivatives of $S$ with respect to $V(y)$. These are related by the chain rule:
\begin{align}
\frac{dS}{d\lambda} &= \int d^{d-2}y~ \frac{\delta S}{\delta V(y)}\dot V(y),\\
\frac{d^2S}{d\lambda^2} &= \int d^{d-2}yd^{d-2}y'~ \frac{\delta^2 S}{\delta V(y)\delta V(y')}\dot V(y) \dot V(y').
\end{align}
We can parametrize the second functional derivative as follows:
\be
\frac{\delta^2 S}{\delta V(y)\delta V(y')} = S_{vv}''(y) \delta^{(d-2)}(y-y') + \left(\frac{\delta^2 S}{\delta V(y)\delta V(y')}\right)_{\rm od}.
\ee
We have extracted the $\delta$-function term explicitly, which we sometimes refer to as the ``diagonal" part, and the remainder carries the label ``od" for ``off-diagonal." Note that the off-diagonal part of the variation does not have to vanish at $y=y'$. The quantity $S_{vv}''$ is the same as $S''$ in~\cite{Bousso:2016aa, Koeller:2016aa,Bousso:2015mna}.

In addition to the entropy of the region $\mathcal{R}$, we can define the vacuum-subtracted modular energy, $\Delta \ev{K}$, and relative entropy with respect to the vacuum, $S_{\rm rel}$, associated to the region $\mathcal{R}$. The modular energy is given by the boost energy along each generator of the null plane~\cite{Casini:2017aa}:
\be\label{eq-modham}
\Delta \ev{K} = 2\pi \int d^{d-2}y \int_{V(y)}^\infty dv~(v - V(y)) \ev{T_{vv}}.
\ee
The relative entropy is defined as the difference between the vacuum-subtracted modular energy and the vacuum-subtracted entropy:
\be
S_{\rm rel} = \Delta \ev{K} - \Delta S.
\ee
For the regions we are talking about, the entropy of the vacuum is stationary and so drops out when we take derivatives of $S_{\rm rel}$. Then for a one-parameter family of regions we have the relations
\begin{align}
\frac{dS_{\rm rel}}{d\lambda} &= -\int d^{d-2}y \left[\frac{\delta S}{\delta V(y)}+2\pi\int_{V(y)}^\infty dv~ \ev{T_{vv}}\right]\dot V(y),\label{eq-entderiv}\\
\frac{d^2S_{\rm rel}}{d\lambda^2} &=  \int d^{d-2}y ~ \left(2\pi \ev{T_{vv}}- S_{vv}''\right)\dot V(y)^2 -\int d^{d-2}yd^{d-2}y'~ \left(\frac{\delta^2 S}{\delta V(y)\delta V(y')}\right)_{\rm od}\dot V(y) \dot V(y').
\end{align}
Note here that our conjectured equation \eqref{eq-nullconjecture} can be restated as saying that the diagonal second variation of the relative entropy is zero. These equations will be mirrored holographically in Section~\ref{sec:bulkpert} below.


\subsection{The Bulk Setup}\label{sec-bulksetup}
While we have a few remarks on the free-field and weakly-interacting cases in Appendix~\ref{sec-free}, most of our nontrivial evidence for \eqref{eq-nullconjecture} and \eqref{eq-conjecture} comes from holography. In this section we will describe the holographic setup for the calculations outlined above. We are actually able to do without much of this machinery in Section~\ref{sec:bulkpert}, though it will become important afterward.

The boundary theory is a quantum field theory in $d$-dimensional Minkowski space obtained by deforming a CFT with relevant couplings. We take the bulk metric to be in Fefferman-Graham gauge (at least near the boundary) and choose to set the AdS length to one:
\be\label{eq-bulkmetric}
ds_{d+1}^2 = \frac{1}{z^2} \left(dz^2 - 2 dudv + d\vec{y}_{d-2}^2 + \gamma_{\mu \nu} dx^{\mu} dx^{\nu}\right).
\ee
Here $x^\mu$ stands for $u$, $v$, or $y$. In the small-$z$ expansion, the metric $\gamma_{\mu \nu}$ is given by~\cite{Hung:2011ta}\footnote{For the purposes of this discussion, we will assume all operators have generic scaling dimensions. In the generic case on a flat background a $\log z$ term in the metric expansion is unnecessary.}
\be\label{eq-smallzmetric}
\gamma_{\mu \nu} = \sum_{\alpha} \gamma^{(\alpha)}_{\mu \nu}z^{\alpha}
\ee
In a fully-quantum treatment, $\gamma_{\mu\nu}$ is an operator in the bulk theory and we would need to take the expectation value of any geometric expression to extract a numerical result. Then there would be a difference between, say, $\ev{\gamma_{\mu\nu}}^2$ and $\ev{\gamma_{\mu\nu}^2}$ that we would have to resolve in order to move beyond leading order in a semiclassical expansion. A consequence of our analysis below is that only expressions which are linear $\gamma_{\mu\nu}$ end up being important for proving \eqref{eq-nullconjecture} and \eqref{eq-conjecture}, and thus this potential difficulty is avoided. With that in mind, we will treat the bulk geometry as classical for ease of presentation.

The term at order $z^d$ in \eqref{eq-smallzmetric}, $\gamma_{\mu\nu}^{(d)}$, contains information about $\ev{T_{\mu \nu}}$~\cite{deHaro:2000vlm}. We will review the dictionary below. The terms at lower orders than $z^d$ are associated with low-dimension operators in the theory~\cite{Hung:2011ta}. If $\mathcal{O}$ is a relevant operator of dimension $\Delta$ and coupling $g$, then possible such terms that we need to be aware of include
\be\label{eq-fterms}
\ev{\mathcal{O}^m} \eta_{\mu\nu}z^{m\Delta},~~~~g^m \eta_{\mu\nu}z^{m(d-\Delta)},~~~~g\ev{\mathcal{O}}\eta_{\mu\nu}z^d,
\ee
with $m\geq 2$. The coupling $g$, when present, is a constant. With only a single operator, terms involving derivatives of $\mathcal{O}$ will always be of higher order than $z^d$ as long as the unitarity bound $\Delta >(d-2)/2$ is obeyed. When there is more than one low-dimension operator then we can also have terms with different combinatorial mixes of couplings and expectation values~\cite{Marolf:2016aa}. In this case, there could also be terms of the form
\be\label{eq-sdterms}
g_1^l\ev{\mathcal{O}_2}\eta_{\mu\nu}z^{l(d-\Delta_1)+\Delta_2},~~~~ g_1^l \partial_{\mu}\partial_{\nu} \ev{\mathcal{O}_2}z^{l(d-\Delta_1)+\Delta_2+2}
\ee
where $\mathcal{O}_1$ and $\mathcal{O}_2$ are two operators and $g_1$ is a relevant coupling associated to $\mathcal{O}_1$. There are other possibilities as well, but we will not need to enumerate them. In order to demonstrate the cancellation of divergences explicitly in \eqref{eq-nullconjecture}, we would need to make use of certain relationships among the various parts of the small-$z$ expansion of the metric. Since there are general arguments for the finiteness of \eqref{eq-nullconjecture}, we will be content to show that the leading state-dependent divergences cancel.\footnote{In other words, we will only explicitly demonstrate the finiteness of \eqref{eq-nullconjecture} given some conditions on the operator dimensions which make the terms we display the only ones that are around.} To that end, we will need the following fact. Suppose that in the sum \eqref{eq-smallzmetric} there is a term of the form $\gamma^{(\alpha)}_{\mu\nu} = \gamma^{(\alpha)}\eta_{\mu\nu}$. Then, assuming that $\alpha$ cannot be written as $\alpha_1+\alpha_2$ for some other $\alpha_1$, $\alpha_2$ occuring in the sum, there will be another term $\gamma^{(\alpha+2)}_{\mu\nu}$ with a null-null component given by
\be\label{eq-alpha2}
\gamma^{(\alpha+2)}_{vv} = \frac{d-2}{(\alpha+2)(d-2-\alpha)}\partial_v^2\gamma^{(\alpha)}.
\ee
This equation is obtained by solving Einstein's equations at small-$z$~\cite{deHaro:2000vlm, Hung:2011ta}. Four-derivative terms are also possible, at order $\alpha+4$, but if $d\leq 6$ then the unitarity bound ensures that $\alpha + 4 > d$. For simplicity we will ignore those terms in this section, but with a little more effort they can also be accounted for.

\paragraph{Holographic Entropy and its Variations}

Our tool for computing the entropy is the Ryu-Takayanagi holographic entropy formula~\cite{Ryu:2006bv, Hubeny:2007xt} including the first quantum corrections~\cite{Faulkner:2013ana},\footnote{In this section and in our main analysis we are only working to next-to-leading order so that the prescriptions of \cite{Faulkner:2013ana} and \cite{Engelhardt:2014aa,Dong:2017aa} agree. If we wanted to work to higher orders in $1/N$, we would need to use the quantum extremal surface prescription instead~\cite{Engelhardt:2014aa,Dong:2017aa}. We discuss this further in Section~\ref{sec-higherorder}.}
\be\label{eq-RT}
S  = \frac{A_{\rm ext}}{4G_N} + S_{\rm bulk}.
\ee
$A_{\rm ext}$ refers to the area of the extremal area surface anchored to $\partial \mathcal{R}$ at $z=0$. The dictionary for computing variations in the entropy as a function of $V(y)$ was laid out in~\cite{Koeller:2016aa} as follows. Let the bulk location of the extremal surface be given by
\be\label{eq-Xbar}
x^\mu = \bar{X}^\mu(y,z) = X^\mu(y) + z^2 X^\mu_{(2)}(y) +\cdots + z^{d}\log z X^{\mu}_{\text{log}} +z^d X^\mu_{(d)}+\cdots,
\ee
where the log term is important for even dimensions and the in the case of relevant deformations with particular operator dimensions. $X^\mu(y)$ are the embedding functions of $\partial \mathcal{R}$ and $\bar{X}^\mu(y,z)$ satisfies the extremal surface equation,
\be\label{eq-extsurfeqn}
\frac{1}{\sqrt{H}}\partial_\alpha\left(\sqrt{H}H^{\alpha\beta}\partial_\beta \bar{X}^\mu\right) + \Gamma_{\rho\sigma}^\mu H^{\alpha\beta}\partial_\alpha \bar{X}^\rho\partial_\beta \bar{X}^\sigma = 0,
\ee
where $H$ is the induced metric on the extremal surface and $\Gamma$ are bulk Christoffel symbols. Note that we have introduced the notation $\bar{X}^\mu$ for the bulk extremal surface coordinates which approach $X^\mu$ on the boundary. We will be interested in computing $\delta A_{\rm ext}/\delta X^\mu(y)$, which by extremality is a pure boundary term evaluated at a $z=\epsilon$ cutoff surface:
\be\label{eq-delA}
\delta A_{\rm ext} = \delta \int d^{d-2}ydz~\sqrt{H} = - \int_{z = \epsilon} d^{d-2}y ~\sqrt{H} H^{zz}g_{\mu\nu}\partial_z\bar{X}^\mu \delta \bar{X}^\nu.
\ee
All of the factors appearing in the integrand need to be expanded in $\epsilon$. The result will be a power series in $\epsilon$ containing divergent terms as well as finite terms:
\begin{align}\label{eqn:Avar}
\frac{\delta A_{\rm ext}}{\delta X^\mu} = -\frac{K_\mu}{(d-2)\epsilon^{d-2}} + (\text{lower-order divergences in } \epsilon)  - (d\,X^{(d)}_\mu + X^{{(\text{log})}}_{\mu}) + O(\epsilon).
\end{align}
Here $K_\mu$ is the extrinsic curvature of the entangling surface. We need to ensure that all divergences cancel or otherwise vanish in \eqref{eq-nullconjecture} and \eqref{eq-conjecture} in order that these be well-defined statements. So here we will explain the structure of the divergences in the entropy variations, as well as how to extract the finite part.

\paragraph{Null Variations}
First, we will consider the special case $X^\mu(y) = V(y)$, which is the relevant case for \eqref{eq-nullconjecture}. If there are no terms of the form \eqref{eq-sdterms} in the metric, then the situation reduces to that of~\cite{Koeller:2016aa}, in which it was shown that the divergent terms in \eqref{eqn:Avar} are absent as long as the entangling surface $\partial\mathcal{R}$ is locally constrained to lie in a null plane. If there are state-dependent terms of the form \eqref{eq-sdterms} in the metric, then there will be non-vanishing divergent contributions to $\delta A_{\rm ext}/\delta V(y)$ proportional to, e.g., $g_1 \partial_v \ev{\mathcal{O}_2}$. In general, an extra term at order $z^\alpha$ in the metric leads to a contribution at order $\alpha+2$ in $\bar{X}^\mu$ that we can obtain by solving \eqref{eq-extsurfeqn} at small $z$. We only need to concern ourselves with terms that have $\alpha + 2 < d$, as those are the ones which lead to divergences. As mentioned above, for $d \leq 6$ the only terms in the metric at order $\alpha$ such that $\alpha + 2 < d$ are those of the form $\gamma^{(\alpha)}_{\mu\nu} = \gamma^{(\alpha)} \eta_{\mu\nu}$. After solving the extremal surface equation in the presence of such a term we find
\be
(\alpha+2)(\alpha+2-d)X^\mu_{(\alpha+2)} = \frac{2(d-2)-\alpha d}{2(d-2)} K^\mu \gamma^{(\alpha)} + \frac{d-2}{2}\partial^\mu \gamma^{(\alpha)}.
\ee
Plugging this in to \eqref{eq-delA} leads to
\be\label{eq-delA2}
\frac{\delta A_{\rm ext}}{\delta V(y)} = \frac{d-2}{2(d-2-\alpha)\epsilon^{d-2-\alpha}}\partial_v \gamma^{(\alpha)}(y)+ dU_{(d)}(y) + \frac{\delta S_{\text{bulk}}}{\delta V(y)},
\ee
where we have eliminated a potential log term by restricting ourselves to the case of generic operator dimensions. The non-generic case can be recovered later as a limit. Using this, we can find the leading-order contribution to the second variation of the entropy:
\be\label{eq-delta2SUV}
\frac{\delta^2 S}{\delta V(y) \delta V(y')} =  \frac{d-2}{8G_N(d-2-\alpha)\epsilon^{d-2-\alpha}}\partial^2_v \gamma^{(\alpha)}(y)\delta^{(d-2)}(y-y')+ \frac{d}{4G_N} \frac{\delta U_{(d)}(y)}{\delta V(y')} + \frac{\delta^2 S_{\text{bulk}}}{\delta V(y) \delta V(y')}.
\ee
Even though this is a very complicated expression in general, we will be able to extract the $\delta$-function contribution and see that it is given by $\ev{T_{vv}}$ as in \eqref{eq-nullconjecture}.

\paragraph{Non-Null Variations}

For a general non-null variation we lose some of the simplifications present in the non-null case. One additional assumption we will make in Section~\ref{sec:nonnull} is to consider entangling surfaces which are planar prior to being deformed, which simplifies some of the geometric expressions. More importantly, however, notice that \eqref{eq-conjecture} only makes reference to the vacuum-subtracted entropy variation, $\Delta S_{\mu\nu}''$, and not $S_{\mu\nu}''$ itself. So any state-independent terms in \eqref{eqn:Avar} can be ignored. Furthermore, for the discussion of the non-null variations we are only going to consider theories where relevant couplings (if present) have mass dimension greater than $d/2$, and states where operators of dimension $\Delta \leq d/2$ have vanishing expectation values in the vicinity of the entangling surface. The result of these restrictions is that terms like \eqref{eq-sdterms} will not be present in the metric up to order $z^d$, and so there will be no state-dependent entropy divergences. Thus for our analysis of non-null deformations, it follows from \eqref{eqn:Avar} that
\be\label{eq-delta2nonnull}
\frac{\delta^2 \Delta S}{\delta X^\mu(y) \delta X^\nu(y')} =  -\frac{d}{4G_N}\Delta\left( \frac{\delta X_\mu^{(d)}(y)}{\delta X^\nu (y')}\right) + \frac{\delta^2 \Delta S_{\text{bulk}}}{\delta X^{\mu}(y) \delta X^\nu (y')}.
\ee
In Section \ref{sec:nonnull} we will also not deal explicitly with the bulk entropy term, but we expect its contributions to be qualitatively similar to the bulk entropy term in the null case.

\paragraph{Identification of the Stress Tensor}

We will also need a holographic formula for the stress tensor, $\ev{T_{\mu\nu}}$. Normally a renormalization procedure is required to define a finite stress tensor. Since our conjectures \eqref{eq-nullconjecture} and \eqref{eq-conjecture} are meant to be finite equations, it will be enough to regulate the stress tensor with a cutoff as we did with the entropy above.\footnote{We still want to define the stress tensor so that $\ev{T_{\mu\nu}}=0$ in vacuum, so the constant vacuum energy term will be subtracted.}

By definition, the (regulated) stress tensor is computed as the derivative of the regulated action:
\be \label{eq-Tmunu}
\ev{T_{\mu\nu}} = \frac{2}{\sqrt{g}}\frac{\delta I_{\rm reg}}{\delta g^{\mu\nu}}- (\text{vacuum energy})~.
\ee
In holography, the regulated action is defined as the action of the bulk spacetime within the $z=\epsilon$ cutoff surface, plus additional boundary terms (like the Gibbons-Hawking-York term) which are necessary to make the variational principle well-defined.~\cite{deHaro:2000vlm, Klebanov:1999tb}. For Einstein gravity in the bulk with minimally-coupled matter fields, the regulated stress tensor is then given by the Brown-York stress tensor evaluated on the $z=\epsilon$ cutoff surface~\cite{Balasubramanian:aa}:\footnote{Care must be taken to impose the correct boundary conditions at $z=\epsilon$. Since we are interested in a flat-space result, we must place a flat metric boundary condition at $z=\epsilon$ before taking $\epsilon \to 0$. This is the only way to get the divergences to cancel out properly between the entropy and the energy in \eqref{eq-nullconjecture}, and this treatment of the boundary condition is especially important if one wants to extend the analysis to curved space~\cite{Akers:2017aa}.}
\begin{align}\label{eq:BY}
\frac{2}{\sqrt{g}}\frac{\delta I_{\rm reg}}{\delta g^{\mu\nu}} &= \frac{-1}{8\pi G_N\epsilon^{d-2}} \left(K_{\mu\nu} - \frac{1}{2} Kg_{\mu\nu}(x,\epsilon)\right)  \nonumber \\
&= \frac{-1}{8\pi G_N\epsilon^{d-2}} \left(-\frac{1}{2\epsilon}\partial_{\epsilon}\gamma_{\mu\nu}(x,\epsilon) +\frac{1}{2\epsilon}\eta_{\mu\nu}\eta^{\rho\sigma}\partial_\epsilon\gamma_{\rho\sigma}(x,\epsilon) + \frac{1-d}{\epsilon^2} \eta_{\mu\nu}\right)
\end{align}
Any state-dependent terms in the metric that occur at order $z^\alpha$ with $\alpha < d$ will contribute to divergences in the stress tensor. In particular, when we discuss null variations we will find contributions from terms of the form \eqref{eq-alpha2}. In total we find
\begin{align}
\ev{T_{vv}} &=\frac{\alpha+2}{16\pi G_N\epsilon^{d-2-\alpha}}\gamma^{(\alpha+2)}_{vv} + \frac{d}{16\pi G_N} \gamma_{vv}^{(d)}\nonumber\\
&= \frac{d-2}{16\pi G_N(d-2-\alpha)\epsilon^{d-2-\alpha}}\partial^2_v\gamma^{(\alpha)}+ \frac{d}{16\pi G_N} \gamma_{vv}^{(d)}.
\end{align}
In the second line we used \eqref{eq-alpha2}. Comparing this to \eqref{eq-delta2SUV}, we see that the divergences indeed cancel out in \eqref{eq-nullconjecture}.

For the non-null case we have additional difficulties. One can easily see that, in general, there are state-dependent divergences in $\ev{T_{\mu\nu}}$ that do not  appear in $S_{\mu\nu}''$. For example, if there are operators of dimension $\Delta <d/2$ in the theory then there will be a term in $\gamma_{\mu\nu}$ at order $z^{2\Delta}$ proportional to $\ev{\mathcal{O}^2}\eta_{\mu\nu}$. By the unitary bound, $2\Delta >d-2$, such a term will not contribute divergences to $S_{\mu\nu}''$, but it will contribute divergences to the stress-tensor of the form
\begin{align}
\braket{T_{\mu\nu}}|_{\epsilon^{2\Delta-d}} \propto \epsilon^{2\Delta - d}\ev{\mathcal{O}^2}\eta_{\mu\nu}.
\end{align}
Thus, when we derive the relationship \eqref{eq-conjecture} in Section~\ref{sec:nonnull}, we will put sufficient restrictions on the theory and the states in consideration so that both sides of the equality are finite and well-defined. As in the case of the entropy variation, all divergences in $\ev{T_{\mu\nu}}$ can be eliminated by restricting the theory so that any nonzero relevant couplings have mass dimension greater than $d/2$, and by restricting the state so that operators of dimension $\Delta \leq d/2$ have vanishing expectation values (at least locally near the entangling surface). When this is true, the metric perturbation $\gamma_{\mu\nu}$ starts at order $z^d$, and so $\ev{T_{\mu\nu}}$ will be finite. Furthermore, we can treat the stress tensor as being effectively traceless even though we are not in a CFT. That is because in general the trace is proportional to products of couplings and scalar expectation values, $g\ev{\mathcal{O}}$, but with our restrictions on the theory and state there is no pair of nonzero coupling and operator expectation value with total dimension adding up to $d$.  The end result is the standard formula for the stress tensor familiar from holographic renormalization~\cite{deHaro:2000vlm}:
\be\label{eq-CFTstress}
\ev{T_{\mu\nu}} = \frac{d}{16\pi G_N}\gamma_{\mu\nu}^{(d)} .
\ee
We will make use of this formula in Section~\ref{sec:nonnull}.


\section{Null Deformations and Perturbative Geometry}\label{sec:bulkpert}

In this section we will prove the relation $S_{vv}'' = 2\pi \ev{T_{vv}}$ for states with geometries corresponding to perturbations of vacuum AdS where it suffices to work to linear order in the metric perturbation. This includes classical as well as quantum states. Below in Section~\ref{sec:genproof} we will extend our results to non-perturbative geometries.

The arguments presented here can be repeated for linearized perturbations to a non-AdS vacuum, i.e., the vacuum of a non-CFT. We restrict ourselves to the AdS case because explicit solutions to the equations are available, and the AdS case also suffices for nearly all applications in the following sections. We will see in Section~\ref{sec:genproof} that in certain situations an appeal to the non-AdS vacuum case is necessary, but because of general arguments (like the known form of the modular Hamiltonian as discussed in the Introduction) we know that the non-AdS case should not behave differently than the AdS case.


\subsection{Bulk and Boundary Relative Entropies}

In~\cite{Jafferis:2015del} it was argued that bulk and boundary relative entropies are identical:
\be\label{eq-JLMS}
S_{\rm rel} = S_{\rm rel, bulk},
\ee
where $S_{\rm rel, bulk}$ is calculated using the bulk quantum state restricted to the entanglement wedge of the boundary region $\mathcal{R}$ --- the region of the bulk bounded by the extremal surface and $\mathcal{R}$.\footnote{At higher orders in $1/N$ this equation is corrected~\cite{Faulkner:2013ana, Dong:2013qoa, Engelhardt:2014gca}. We will not go into these corrections in detail, but will make a few comments below in Section~\ref{sec-higherorder}.}

We already discussed in Section~\ref{sec-ftsetup} the form of $S_{\rm rel}$ for the regions we are considering, but to leading order in bulk perturbation theory there is an analogous simple formula for $S_{\rm rel, bulk}$. We only need to know two simple facts. First, if $\partial \mathcal{R}$ is restricted to lie in the $u=0$ plane on the boundary then, to leading order, the extremal surface in the bulk also lies in the $u=0$ plane. Second, to leading order the bulk modular energy corresponding to such a region is given by the AdS analogue of \eqref{eq-modham}:
\be\label{eq-DeltaKbulk}
\Delta K_{\rm bulk} = 2\pi \int \frac{dzd^{d-2}y}{z^{d-1}}\int_{\bar{V}(y)}^\infty dv~(v-\bar{V}(y,z))\ev{T_{vv}^{\rm bulk}}.
\ee
In keeping with our earlier notation, $\bar{V}(y,z)$ gives the location of the bulk extremal surface with $\bar{V}(y,z=0) = V(y)$. Now we simply solve \eqref{eq-JLMS} for the vacuum-subtracted boundary entropy $\Delta S$,
\be
\Delta S = \Delta \ev{K} - \Delta \ev{K_{\rm bulk}} + \Delta S_{\rm bulk},
\ee
and take two derivatives with respect to a deformation parameter $\lambda$ to find
\be\label{eq-Sderiv}
\frac{d^2S}{d\lambda^2} = 2\pi \int d^{d-2}y ~\ev{T_{vv}}\dot{V}^2 - 2\pi \int \frac{dzd^{d-2}y}{z^{d-1}}~\ev{T_{vv}^{\rm bulk}}\dot{\bar{V}}^2 + \frac{d^2S_{\rm bulk}}{d\lambda^2}.
\ee
The first term represents a contribution of $2\pi \ev{T_{vv}}$ to $S_{vv}''$. So \eqref{eq-nullconjecture}, $S_{vv}'' = 2\pi \ev{T_{vv}}$, amounts to showing that the remaining two terms do not contribute to $S_{vv}''$. We examine them both in the next section.

\subsection{Proof of the Conjecture}

From the discussion around \eqref{eq-Sderiv}, the conjecture $S_{vv}'' = 2\pi \ev{T_{vv}}$ amounts to the statement that the terms
\be
- 2\pi \int \frac{dzd^{d-2}y}{z^{d-1}}~\ev{T_{vv}^{\rm bulk}}\dot{\bar{V}}^2 + \frac{d^2S_{\rm bulk}}{d\lambda^2}.
\ee
do not contribute a $\delta$-function to the second variation of $S$. Together these terms comprise the second derivative of the bulk relative entropy. We treat the two terms individually.

\paragraph{Bulk Modular Energy}
The modular energy term is simple to evaluate. Note that \eqref{eq-DeltaKbulk} depends on the entangling surface $V(y)$ through the extremal surface $\bar{V}(y,z)$. So functional derivatives of that expression with respect to $V(y)$ involves factors of $\delta \bar{V}(y,z)/\delta V(y')$. This is the boundary-to-bulk propagator of the extremal surface equation in pure AdS. The result, which can be extracted from our discussion in later sections, is~\cite{Nozaki:2013aa}
\be
\frac{\delta \bar{V}(y,z)}{\delta V(y)} = \frac{2^{d-2}\Gamma(\frac{d-1}{2})}{\pi^{\frac{d-1}{2}}} \frac{z^d}{(z^2+(y-y')^2)^{d-1}} . 
\ee
Then we have
\be\label{eqn:Kbulkvar}
\frac{\delta^2 K_{\rm bulk}}{\delta V(y_1)\delta V(y_2)} = 2\pi \left(\frac{2^{d-2}\Gamma(\frac{d-1}{2})}{\pi^{\frac{d-1}{2}}}\right)^2 \int \frac{dzd^{d-2}y}{z^{d-1}}~\ev{T_{vv}^{\rm bulk}}\frac{z^{2d}}{(z^2 + (y-y_1)^2)^{d-1}(z^2 + (y-y_2)^2)^{d-1}}
\ee
We can diagnose the presence of a $\delta$-function by integrating with respect to $y_1$ over a small neighborhood of $y_2$. If the result remains finite as the size of the neighborhood goes to zero, then we have a $\delta$-function. Whether or not this happens depends on the falloff conditions on $\ev{T_{vv}^{\rm bulk}}$ near $z=0$, which in turn depends on the matter content of the bulk theory. If we suppose $\ev{T_{vv}^{\rm bulk}} \sim z^{\beta}$ as $z\to 0$, then it is easy to see that there is no $\delta$-function so long as
\be\label{eq-unitarity}
\beta > d-2.
\ee
For scalar fields in the bulk, $T^{\rm bulk}_{vv} \sim (\partial_v \phi)^2\sim z^{2\Delta}$ where $\Delta$ is the dimension of the dual operator. This is even true when the non-normalizable mode $\phi \sim g z^{d-\Delta}$ is turned on, as long as the coupling $g$ is constant. For bulk Dirac fields, $T^{\rm bulk}_{vv} \sim \bar{\psi} \Gamma_v \nabla_v \psi\sim z^{2\Delta -1}$. In either case, equation \eqref{eq-unitarity} reduces to the unitarity bound on the dual operator dimension, $\Delta > (d-2)/2 +s$, where $s=0, 1/2$ is the spin. In the limiting case where the unitarity bound is saturated and the dual operator is a free scalar or free fermion, one may find a $\delta$-function in \eqref{eqn:Kbulkvar}. Indeed, in Appendix \ref{sec-free} we find extra contributions to $S''_{vv}$ besides $2\pi \ev{T_{vv}}$ for a free scalar field, so the appearance of an additional $\delta$-function in this case is an expected feature. The case of a free fermion has not yet been worked out in the field theory, but methods similar to those in Appendix \ref{sec-free} should be applicable. For operators which do not saturate the unitarity bound, we have shown that $\Delta K_{\rm bulk}$ does not contribute to $S''_{vv}$.

\paragraph{Bulk Entropy}
It is much more difficult to make statements about $d^2 S_{\rm bulk}/d\lambda^2$. In a coherent bulk state we know that $d^2S_{\rm bulk}/d\lambda^2=0$, so for that class of states we are done.\footnote{In this section we treat the bulk matter fields as free. If we turn on weak interactions, then the comments of Appendix~\ref{sec-weak} apply. Qualitatively nothing changes.} More generally, we can write
\begin{align}\label{eq-sbulkvar}
&\frac{\delta^2 S_{\rm bulk}}{\delta V(y_1)\delta V(y_2)}  = \nonumber \\ &\left(\frac{2^{d-2}\Gamma(\frac{d-1}{2})}{\pi^{\frac{d-1}{2}}}\right)^2\int d^{d-2}ydzd^{d-2}y'dz' \frac{\delta^2 S_{\rm bulk}}{\delta \bar{V}(y,z)\delta \bar{V}(y',z')} \frac{(zz')^d}{(z^2 + (y-y_1)^2)^{d-1}({z'}^2 + (y'-y_2)^2)^{d-1}}
\end{align}
and ask what sort of behavior would be required of $\delta^2 S_{\rm bulk}/\delta \bar{V}(y,z)\bar{V}(y',z')$ in order to lead to a $\delta$-function in $y_1-y_2$.

As a toy model, we can imagine a collection of particles on the $u=0$ surface which are entangled in a way that depends on their distance from each other. This is a fairly general ansatz for the state of a free theory in the formalism of null quantization~\cite{Wall:2011hj}. At small $z$ (which is the dominant part for our calculation) this would correspond to a second variation of the form
\be\label{eq-bulkentform}
\frac{\delta^2 S_{\rm bulk}}{\delta \bar{V}(y,z)\delta \bar{V}(y',z')} \sim \frac{ (zz')^\Delta}{(zz')^{d-1}} F\left(\frac{zz'}{(z-z')^2 + (y-y')^2}\right).
\ee
The factor $(zz')^\Delta/(zz')^{d-1}$ reflects that entropy variations should be proportional to the amount of matter present at locations $z$ and $z'$. The numerator encodes the falloff conditions on the density of particles in a way that is consistent with the falloff conditions for a bosonic matter field, and the denominator is a measure factor that converts coordinate areas to physical areas. The function $F$ is arbitrary.

With the assumption of \eqref{eq-bulkentform}, a constant rescaling of all coordinates by $\alpha$ leads to an overall factor of $\alpha^{4-2d+2\Delta}$ in \eqref{eq-sbulkvar}. A $\delta$-function in $y_1-y_2$ would scale like $\alpha^{2-d}$, and anything that scales with a power of $\alpha$ less than $2-d$ would correspond to a more-divergent distribution, like the derivative of a $\delta$-function. As long as $\Delta > (d-2)/2$ this is avoided, and a $\delta$-function is only present when the unitarity bound $\Delta = (d-2)/2$ is saturated. This is consistent with what we found previously for the modular energy, and with our general expectations for free theories.

\section{Non-Perturbative Bulk Geometry}\label{sec:genproof}

Now we turn to a proof that applies for a general bulk geometry, still restricting the deformations to be null on the boundary. We will use the techniques outlined in Section~\ref{sec-bulksetup}, which relate the entropy variations to changes in the bulk extremal surface location. At first we will stick to boundary regions where  $\partial \mathcal{R}$ is restricted to a null plane, leaving a generalization to regions where $\partial \mathcal{R}$ only satisfies certain local conditions for Section \ref{sec-localconditions}.

\subsection{Extremal Surface Equations}

\begin{figure}
	\centering
	\includegraphics[scale=.7]{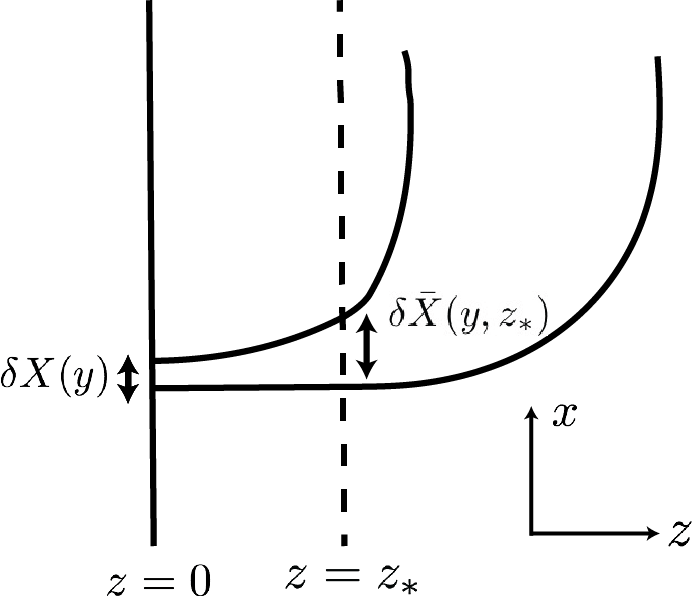}
	\caption{By restricting attention to $z<z_{*}$ the geometry is close to pure AdS, and we can solve for $\delta \bar{X}$ perturbatively. All of the $z<z_{*}$ data imprints itself as boundary conditions at $z=z_{*}$. We show that these boundary conditions are unimportant for our analysis, which means that a perturbative calculation is enough.}\label{fig:zbar}
\end{figure}

\paragraph{Small $z$, Large $k$}

The extremal surface equation \eqref{eq-extsurfeqn} for $\bar{U}$ and $\bar{V}$ is a very complicated equation. If we perturb the boundary conditions by taking $V\to V + \delta V$, then the responses $\delta \bar{U}$ and $\delta \bar{V}$ will satisfy the linearized extremal surface equation, which is a bit simpler. It may be that the coordinates we have chosen are not well-suited to describing the surface perturbations deep into the bulk. That problem is solved by only aiming to analyze the equations in the range $z < z_*$ for some small but finite $z_*$. In fact, by choosing $z_*$ small enough we can say that the spacetime is perturbatively close to vacuum AdS, with the perturbation given by the Fefferman-Graham expansion \eqref{eq-smallzmetric}. Since the corrections to the vacuum geometry are small when $z_*$ is small, the extremal surface equation reduces to the vacuum extremal surface equation plus perturbative corrections. All of the deep-in-the-bulk physics is encoded in boundary conditions at $z=z_*$. The situation is illustrated in Fig.~\ref{fig:zbar}

The boundary conditions at $z=z_*$ are essentially impossible to find in the general case, so the restriction to $z<z_*$ does not make the problem of finding the extremal surface any easier. However, according to \eqref{eq-delta2SUV} all we are interested in is the $\delta$-function part of $\delta U_{(d)}$. It will turn out that this quantity is actually independent of those boundary conditions. 

The idea is very simple. In Fourier space a $\delta$-function has constant magnitude. That means it does not go to zero at large values of $k$, unlike the Fourier transform of a smooth function. So the strategy will be to analyze the extremal surface equation in Fourier space at large $k$. We will see that the large-$k$ response of $\bar{U}$ (and hence $U_{(d)}$) is completely determined by near-boundary physics, and in particular will match the results we found in previous sections. This will establish that $S_{vv}''= 2\pi \ev{T_{vv}}$ for very general bulk states.

\paragraph{Integral Equation for $\bar{U}$}

We will begin by finding an integral equation for $\bar{U}$ in the range $z < z_*$. Since $\bar{U}$ vanishes at $z=0$ it must remain small throughout $z<z_*$, as long as $z_*$ is small enough, and so we can use perturbation theory to find $\bar{U}$ in that range. Then we will compute the response of $\bar{U}$ to variations of the boundary conditions $V$ at $z=0$. Expanding (\ref{eq-extsurfeqn}) in small $z$, we can write the equation for $\bar{U}$ as
\be\label{eq-nonlineareq}
\partial^2_a \bar{U}+\partial^2_z \bar{U}+\frac{1-d}{z}\partial_z \bar{U} = J[\gamma_{\mu\nu},\bar{V}, \bar{U}],
\ee
where $\gamma_{\mu\nu}/z^2$ is the deviation of the metric from vacuum AdS, as in \eqref{eq-smallzmetric}. To solve this equation perturbatively we require a Green's function $G(z,y|z',y')$ of the linearized extremal surface equation that vanishes when $z=0$ or $z=z_*$. Then the solution to \eqref{eq-nonlineareq} can be written as 
\be\label{eq-Unonpert}
\bar{U}(y,z) = \int \frac{d^{d-2}y'}{z_*^{d-1}}~ \partial_{z'}G(y,z|y',z_*) \bar{U}(y',z_*)   +  \int_{z<z_*} \frac{d^{d-2}y'dz'}{{z'}^{d-1}}G(y,z|y',z')J(y',z')
\ee
It is important to remember that $J(y,z)$ is itself a functional of $\bar{U}$, and the usual methods of perturbation theory would involve solving for $\bar{U}$ iteratively. It will be more useful for us to look at the Fourier transform of this equation:
\be\label{eq-Unonpertk}
\bar{U}(k,z) = z_*^{1-d}\partial_{z'}G_k(z|z_*) \bar{U}(k,z_*)   +  \int_0^{z_*} \frac{dz'}{{z'}^{d-1}}G_k(z|z')J(k,z').
\ee
The Green's function with the correct boundary conditions is easily obtained from the standard Green's function $G^{\rm AdS}$ by adding a particular solution of the vacuum extremal surface equation. In Fourier space, the answer is
\begin{align}
G_k(z|z') = G^{\rm AdS}_k(z|z') + (zz')^{d/2} I_{d/2}(kz)I_{d/2}(kz')\frac{K_{d/2}(kz_*)}{I_{d/2}(kz_*)}
\end{align}
where
\be\label{eqn:greens}
G^{\rm AdS}_k(z|z') = -\begin{cases}
(zz')^{d/2}  I_{d/2}(kz)K_{d/2}(kz'), & z < z',  \\
(zz')^{d/2} I_{d/2}(kz')K_{d/2}(kz),  & z > z'.
\end{cases}
\ee
In the limit of large $k$, the first term of (\ref{eq-Unonpertk}) becomes exponentially suppressed. So we see that the boundary conditions at $z=z_*$ do not matter. Furthermore, the integration range $z' \gtrsim 1/k$ in the second term also becomes exponentially suppressed. So only the small-$z$ part of the source $J$ contributes at leading order in the large-$k$ limit.

\subsection{Terms in the Source}

Let us consider the form of the source in position space in more detail. We know that $J= J[\bar{U}, \bar{V}, \gamma]$ is a functional of the extremal surface coordinates and the metric perturbation.  We can treat $J$ as a double power series in $\gamma$ and $\bar{U}$ since we are doing perturbation theory in those two parameters. We will repeatedly take advantage of the ``boost" symmetry of the equation: under the coordinate transformation $u \to \alpha u$, $v\to \alpha^{-1}v$, the source must transform as $J\to \alpha J$ in order for the whole equation to be covariant. Since every occurrence of $\bar{V}$ must be accompanied by either a $\gamma$ or $\bar{U}$ to preserve the boost symmetry, $J[\bar{U}, \bar{V}, \gamma]$ is actually a triple power series in all three of its parameters. Another important fact is dimensional analysis, which comes from scaling all coordinates together: $J$ has length dimension $-1$, while $\bar{U}$ and $\bar{V}$ have dimension $1$ and $\gamma$ has dimension zero. This will also be used to restrict the types of terms we can find.

The variation  $\delta \bar{U}$ satisfies an integral equation similar to that of  $\bar{U}$ except with the source, $J$, replaced by the variation of the source, $\delta J$. Like $J$, we can treat $\delta J$ as a power series. Each term in the $\delta J$ power series contains a single $\delta \bar{U}$, $\delta \gamma$, or $\delta \bar{V}$, multiplied by some number of $\bar{U}$, $\bar{V}$, and $\gamma$ factors (and their derivatives).  It is important to note that these unvaried $\bar{U}$, $\bar{V}$, and $\gamma$ factors are smooth, and therefore their Fourier transforms decay at large $k$. So the Fourier transform of a term in $\delta J$ looks schematically like
\be\label{eq-deltaJschematic}
\delta J(k) \sim \int_{k' <<k} dk' ~ h(k')\delta \Psi(k-k'),
\ee
where $\Psi$ is either $\gamma$, $\bar{V}$, $\bar{U}$, or their derivatives and $h$ is the Fourier transform of a smooth function. The $k$-dependence at large $k$ of a given term in $\delta J$ is completely determined by the factor $\delta \Psi$ being varied. The case where $\Psi=\gamma$ can be reduced immediately to the other two, because $\delta \gamma =\delta \bar{V} \partial_v\gamma +  \delta \bar{U}\partial_u\gamma$.

In Fourier space, we can write $\delta J(k,z)$ as a sum of terms of the form $\delta J_{mn}z^{m}k^n$ at small $z$ and large $k$.\footnote{There may also be terms in the source of the form $z^m \log (z)$. Qualitatively these terms behave similarly to the $z^m$ terms as far as the $\delta$-function part of the entropy variation is concerned, so we will not explicitly keep track of them.} Since the effect of $z_*$ is exponentially suppressed at large $k$, we can drop the first term in \eqref{eq-Unonpertk} and push the limit in the second term off to infinity. Additionally, the difference between $G_k(z|z')$ and $G_k^{\rm AdS}(z|z')$ is exponentially suppressed. Thus for our purposes we have
\begin{align}\label{eq:deltaU}
&\delta \bU(k,z) = \sum_{m,n} \int_0^{\infty}G^{\rm AdS}_{k}(z|z')\delta J_{mn}z^mk^n + O(e^{-kz_*}) \\ \nonumber
& = \sum_{m,n}\delta J_{mn}\left(\frac{ k^n z^{2+m}(d-2(m+2))}{d(m+2)(d-m-2)} -z^d 2^{m-d}k^{n-m-2+d}\frac{\Gamma \left(1+\frac{m}{2}\right)\Gamma\left(\frac{m-d+2}{2}\right)}{\Gamma(1+d/2)} \right)+ \mathcal{O}(z^{d+1})
\end{align}
If $m < d-2$ then the first term in \eqref{eq:deltaU} represents a contribution to $\bar{U}$ that could have been obtained by doing the small-$z$ expansion of the extremal surface equation. In a CFT these would consist only of geometric terms that depend on extrinsic curvatures of the entangling surface, but our boundary condition $U=0$ guarantees that those vanish. Still, when a relevant deformation is turned on there may be terms proportional to $g_1^l\partial_v \ev{\mathcal{O}_2}$ which enter $\bar{U}$ at low orders in $z$. An important fact, enforced by the unitarity bound, is that these low-order terms are all linear in expectation values. When $m=d-2$ each of the terms in \eqref{eq:deltaU} becomes singular, but actually the combination above remains finite and generates at $z^d\log z$ term. Since \eqref{eq:deltaU} is well-behaved in this limit, we can treat the non-generic case $m=d-2$ as a limiting case of generic $m$. Thus throughout our discussion below $m$ is assumed to be generic. Finally, for $d>6$ another term proportional to $z^{4+m}$ (and $z^{6+m}$ in $d >8$, etc.) should be included, but for simplicity we have not written it down. Qualitatively it has the same properties as the $z^{2+m}$ term.

Our focus is on the $z^d$ term, as this is where the finite contributions to the entropy variation come from, as in \eqref{eq-delta2SUV}. From \eqref{eq:deltaU}, we see that the $\delta$-function is determined by source terms with $n-m = 2-d$, which corresponds to $k^0$ behavior at large $k$. So our task is simply to enumerate the possible terms in $\delta J$ which have this behavior. We will see that such terms are completely accounted for by the linearized analysis of the previous section,\footnote{As mentioned in the previous section, for simplicity of presentation we are performing our perturbation theory around empty AdS, whereas in complete generality one would want to perform the analysis based around the vacuum of the theory in question. The difference is that some terms which are linear in expectation values $\ev{\mathcal{O}}$ might appear at higher orders in perturbation theory around empty AdS even though they are fully accounted for in the linearized analysis about the correct vacuum.} which completes the proof.

\paragraph{Ingredients}
Before diving into the terms of the source, we will collect all of the facts we need about the function $\bar{U}$, $\bar{V}$, $\gamma$, and their variations. In particular, we will need to know what powers of $k$ and $z$ we can expect them to contribute to the source.

We begin with $\bar{V}$. Unlike $\bar{U}$, $\bar{V}$ does not have any particular boundary condition at $z=0$. Thus the Fefferman-Graham expansion for $\bar{V}$ contains low powers of $z$ that depend on geometric data of the entangling surface. In particular, the boundary condition itself enters $\bar{V}$ at order $z^0$, which is neutral in terms of the $n-m$ counting. That same behavior extends to the variation $\delta \bar{V}$: in Fourier space, the state-independent parts of $\delta \bar{V}$ are functions of the combination $kz$. In other words, we find schematically
\be\label{eq-vvar}
\delta \bar{V} \sim (1 + k^2z^2 + k^4z^4 + \cdots) \delta V.
\ee
The boundary condition $\delta V$ itself is taken to go like $k^0$ at large $k$ (i.e., a $\delta$-function variation). So in terms of our power counting, which only depends on $n-m$, these terms are all completely neutral. So a factor of $\delta \bar{V}$ in the source is ``free'' as far as the power counting is concerned. There will be other terms in $\delta \bar{V}$, even at low powers of $z$, but the terms in \eqref{eq-vvar} are the ones which dominate the $n-m$ counting.

$\bar{U}$ is also an extremal surface coordinate, but it has the restricted boundary condition $U=0$. That means it does not possess terms like those in \eqref{eq-vvar}. The lowest-order-in-$z$ terms that can be present are of the form $g_1^l \partial_v\ev{\mathcal{O}_2}z^{2+l(d-\Delta_1)+ \Delta_2}$. It is only terms like this which contain a single factor of $\mathcal{O}$ that can show up at lower orders than $z^d$, because of the unitarity bound $\Delta > (d-2)/2$. Taking a variation, we find a term in $\delta \bar{U}$ of the form
\be\label{eq-uvar}
\delta \bar{U} \sim g_1^l \partial^2_v\ev{\mathcal{O}_2}\delta V z^{2+l(d-\Delta_1))+ \Delta_2},
\ee
which has $n-m = -(2+l(d-\Delta_1)+ \Delta_2)$.

The final ingredient is the metric perturbation $\gamma$. We don't have to consider variations of $\gamma$ directly, since they can be re-expressed in term s of variations of $\bar{U}$ and $\bar{V}$. $\gamma$ itself has a Fefferman-Graham expansion which in includes information about the stress tensor at order $z^d$, but can have lower-order terms as well that depend on couplings and expectation values of operators. We will see that the important terms in the source that affect the $\delta$-function response are those which are linear in $\gamma$.

\paragraph{Terms with $\delta \bar{U}$}
Now we will analyze the possible terms in the source which can be obtained by piecing together the above ingredients. We begin with terms proportional to $\delta \bar{U}$. As stated above, there are dominant contributions to $\bar{U}$ in terms of the $n-m$ counting which are proportional to derivatives of expectation values of operators.

But $\bar{U}$ does not occur alone in the source $J$: since all terms with $\bar{U}$ alone in the equation of motion are part of the linearized equation of motion on the left-hand-side of \eqref{eq-nonlineareq}. An additional factor of $\bar{V}$ does not affect the dominant $n-m$ value of the term, but the combination $\bar{U}\bar{V}$ is also prevented from appearing in $J$ by boost symmetry. We need to have at least another factor of $\bar{U}$, or else a factor of $\gamma$. The dominant possibility without using $\gamma$ is something of the form $\partial\bar{U}\partial\bar{V}\partial^2\delta \bar{U}$, where derivatives have been inserted to enforce the correct total dimensionality. Taking into account the derivatives, a term like this can have at most $n-m = 3-2(2+l(d-\Delta_1)+ \Delta_2) < 1-d-2l(d-\Delta_1) < 2-d$, using the unitarity bound. So this sort of term will not matter for the $\delta$-function response.

Making use of $\gamma$ allows for more possibilities. Terms of the schematic form $\gamma \delta \bar{U}$ in the source can have $n-m > 2-d$, and if we allow fine-tuning of operator dimensions we can even reach $n-m = 2-d$. These sources are obtained by taking a state-independent term in $\gamma$ which is proportional to some power of $g_1$ and a term in $\delta \bar{U}$ which is proportional to $\partial_v^2\ev{\mathcal{O}_2}$. We can even multiply by more factors of $\gamma$, giving $\gamma^l \delta \bar{U}$ schematically, as well as factors of $\bar{V}$, as long as we don't involve more factors of $\bar{U}$. A second factor of $\bar{U}$ brings with it a large $z$-scaling, so we run into the same problem we had above in the $\bar{U}\bar{V}\delta\bar{U}$ case. The end result is that all of the potentially-important terms in this analysis are linear in the expectation value $\ev{\mathcal{O}}$. That means they are subject to restrictions on the modular Hamiltonian as mentioned in the Introduction, which means that they will actually not show up in \eqref{eq-nullconjecture} despite being allowed by dimensional analysis.

\paragraph{Terms with $\delta \bar{V}$}

Now we consider terms in $\delta J$ that are proportional to a variation $\delta \bar{V}$. As discussed above, $\delta \bar{V}$ has several state-independent terms which are neutral in the $n-m$ counting. Due to the boost symmetry, $\delta \bar{V}$ cannot occur alone in $\delta J$. It must be accompanied by at least two factors of $\bar{U}$ or one factor of $\gamma$. We have already discussed how two factors of $\bar{U}$ have a large-enough $z$-scaling to make the term uninteresting, so it remains to consider factors of $\gamma$.

Terms in the source proportional to $\delta \bar{V}$ with only a single factor of $\gamma$ are those present in the theory of linearized gravity about vacuum AdS. Furthermore, since we argued that boundary conditions at $z=z_*$ do not affect the answer, the Green's function we use to compute the effects of the source is {\em also} the same as we would use in linearized gravity about vacuum AdS. We already considered the linearized gravity setup in Section~\ref{sec:bulkpert}, even though we didn't solve it using the methods of this section. In Section~\ref{sec:bulkpert} we saw that $S''_{vv} = 2\pi \ev{T_{vv}}$, and so it is enough for us now to prove that the general computation of the $\delta$-function terms reduces to the linearized gravity case. There is only one more loose end to consider: terms in $\delta J$ proportional to $\delta \bar{V}$ that have more than one factor of $\gamma$.

With more than a single factor of $\gamma$, it is clear that the only contributions that could possibly be important at large $k$ are those coming from the powers of $z$ less than $z^d$ in \eqref{eq-smallzmetric}. These terms are made up of  couplings $g$, operator expectation values $\ev{\mathcal{O}}$, and their derivatives. In order to have the correct boost scaling, we need to include $v$-derivatives acting on operator expectation values. As we have discussed many times, the unitarity bound prevents any term with more than one factor of $\ev{\mathcal{O}}$ from being important. So just as with the $\delta \bar{U}$ terms discussed previously, all of these terms are subject to constraints from the modular Hamiltonian and hence do not appear in \eqref{eq-nullconjecture}

Our analysis so far has been very simple , but we have reached an important conclusion that bears repeating: the source terms which give the $k^0$ behavior for $\delta U_{(d)}$ were already present in the linearized gravity calculation of the previous section, and we are allowed to use the ordinary Green's function $G^{\rm AdS}$ to compute their effects. In other words, for the purpose of calculating the $\delta$-function response we have reduced the problem to linearized gravity. We have shown previously that the linearized gravity setup leads to $S_{vv}'' = 2\pi \ev{T_{vv}}$, and so our proof is complete.


\section{Non-Null Deformations}\label{sec:nonnull}

Having established $S_{vv}'' = 2\pi \ev{T_{vv}}$ for deformations of entangling surfaces restricted to lie in the plane $u=0$, we will now analyze arbitrary deformations of the entangling surface to prove \eqref{eq-conjecture}. The technique is very similar to that of the previous section. As discussed in Sec \ref{sec-bulksetup}, there are additional assumptions and restrictions we make in this case to help us deal with divergences and to simplify the analysis. First, we restrict attention to theories where all relevant couplings, if present, have mass dimension greater than $d/2$. Second, we restrict the state so that operators with scaling dimension $\Delta \leq d/2$ have vanishing expectation value near the entangling surface. Finally, we restrict the entangling surface itself to be planar prior to taking any variations.

\subsection{New Boundary Conditions}

Above we analyzed deformations within the null plane $u=0$ at small $z$ and large $k$. These limits allowed us to show that the perturbation theory for $\delta U_{(d)}$ reduced to linearized gravity, which we had already studied in Section~\ref{sec:bulkpert}. There strategy here is the same, except we want to be able to perform perturbation theory on both $\bar{U}$ and $\bar{V}$ in order to get more than just the null-null variations. The simplest case, which is all that we will analyze in this work, is to start with the boundary condition $V=0$ at $z=0$ in addition to $U=0$. In other words, we take our undeformed entangling surface to be the $v=u=0$ plane. That is a severe restriction on the type of surface we are considering, but we gain the flexibility of being able to do perturbation theory in both $\bar{U}$ and $\bar{V}$. From \eqref{eq-delta2nonnull},
\be
\frac{\delta^2 \Delta S}{\delta X^\mu(y) \delta X^\nu(y')} =  -\frac{d}{4G_N}\Delta\left( \frac{\delta X_\mu^{(d)}(y)}{\delta X^\nu (y')}\right) + \frac{\delta^2 \Delta S_{\text{bulk}}}{\delta X^\mu(y) \delta X^\nu(y')},
\ee
where $\Delta S$ refers to the vacuum-subtracted entropy. Vacuum subtraction removes all state-independent terms from the entropy, including divergences.\ blue{For the remainder of the section, we will drop the bulk entropy contribution.}

With the $U=V=0$ boundary conditions, we can again write down our perturbative extremal surface equation for the $z<z_*$ part of the bulk. Since the null direction is no longer preferred, we will use a covariant form of the linearized equation:
\be
\partial_a^2 \bar{X}^\mu + \partial^2_z \bar{X}^\mu + \frac{1-d}{z}\partial_z \bar{X}^\mu = J^\mu[\gamma, \bar{X}]
\ee
Following the same steps as in the previous section, we can use Green's functions to solve this equation in Fourier space. There is one new ingredient that we did not have before. When we computed the variation of $U_{(d)}$ with respect to $V$, we were changing the boundary conditions of $\bar{V}$ and computing the response in $\bar{U}$. In particular, the boundary condition of $\bar{U}$ itself remained zero. In the more general setup of this section, we need to compute the response of a particular component of $\bar{X}^\mu$ when its own boundary conditions at $z=0$ are varied.

Since we only care about the $\delta$-function contribution to the entropy variation, we will immediately use $\delta X^\mu(k) = e^{iky_0} \xi^\mu$ as the boundary condition for $\delta \bar{X}^\mu$. Here $\xi^\mu$ is just a constant vector which tells us the direction of the perturbation. The presence of this boundary condition at $z=0$ is simple to account for with one additional term in the integral equation for $\bar{X}^\mu$ compared to \eqref{eq-Unonpertk} in the previous section. In total, we now have
\begin{align}\label{eq-Xmud}
\delta X^{\mu}(k,z) &= z^{d/2}K_{d/2}(kz) \frac{dk^{d/2}}{2^{d/2} \Gamma(1+d/2)} \xi^{\mu} e^{iky_0} \nonumber\\  &+ z_*^{1-d}\partial_{z'}G(z|z_*)\delta \bar{X}^{\mu}(k,z_*) + \int_0^{z_*} \frac{dz'}{z'^{d-1}} G_k(z|z')\delta J^{\mu}(k,z')
\end{align}
As above, in the large-$k$ limit the term coming from boundary conditions at $z=z_*$ (the first term in the second line of \eqref{eq-Xmud}) will drop out and so can be ignored completely. The term from boundary conditions at $z=0$ (the first line of \eqref{eq-Xmud}) will not drop out automatically, and so will contribute to the second entropy variation. This contribution to the entropy variation is known as the entanglement density in the literature and was previously computed in~\cite{Faulkner:2016aa, Bhattacharya:2014vja}. From \eqref{eq-Xmud} it is clear that the entanglement density is completely determined by the AdS Green's function and is therefore state-independent. By restricting attention to the vacuum-subtracted entropy the entanglement density will drop out, and in any case is not proportional to a $\delta$-function.

\subsection{Terms in the Source}

As in the null deformation discussion of Section~\ref{sec:genproof},  we need to compute the effects of the source $\delta J^\mu$. As we did there, we will accomplish this by cataloging the various terms which can appear in the power series expansion of $J^\mu$ as a function of $\bar{X}$ and $\gamma$. Again, terms which scale like $k^nz^m$ ultimately lead to $k^{n-m+d-2}$ dependence at large $k$ for $\delta X_{(d)}^\mu$. Any term in $\delta J^\mu$ will look like $\delta \bar{X}^\nu$ multiplied by some function of $\gamma$ and $\bar{X}$. For the purposes of computing $\delta J^\mu$ only the state-independent parts of $\delta \bar{X}^\nu$, represented by the first line of \eqref{eq-Xmud}, will matter. That is because these terms are a function of the combination $kz$, which means they have $n-m=0$. Now we just have to consider all of the possible combinations of $\gamma$ and $\bar{X}$ which multiply $\delta \bar{X}$. 

There cannot be any terms in $\delta J^\mu$ that are schematically of the form $\bar{X}\delta \bar{X}$ with some derivatives but no factors of $\gamma$. Such a term would have to come from nonlinearities in the vacuum AdS extremal surface equation. That equation is invariant under $\bar{X} \to -\bar{X}$, so all terms have to have odd parity like the linear terms. Anything of the form $\bar{X}\bar{X}\delta \bar{X}$, or higher powers of $\bar{X}$, will not contribute at large $k$ because of power counting: The vanishing boundary condition means that $\bar{X}$ starts at order $z^d$, which means that the most favorable possible term of this type, $(\partial_z \bar{X})^2 \partial_z^2\delta \bar{X}$, still only amounts to a contribution to the entropy variation which scales like $k^{2-d}$.

Now we consider terms which have at least one factor of $\gamma$. Because we have assumed that all couplings have dimension greater than $d/2$ and that expectation values of operators with dimension $\Delta \leq d/2$ vanish, the leading order piece of $\gamma$ scales like $z^d$. Thus we can get contributions to $\delta X_{(d)}$ which go like $k^0$ from source terms which are schematically of the form $\gamma \partial^2\delta\bar{X}$, as well as other combinations. Given their importance, we will analyze terms of the form $\gamma \delta\bar{X}$ below in more detail.

Terms with additional factors of $\bar{X}$ or $\gamma$ beyond the first power of $\gamma$ will not lead to non-decaying behavior at large $k$ because of power counting. So we see that only the linear gravitational backreaction is necessary to completely characterize $\Delta S_{\mu\nu}''$. We will now calculate those terms explicitly.

\subsection{Linearized Geometry}
We have reduced our task to computing $J^\mu$ to linear order in $\gamma$ and $\bar{X}^\mu$ (the latter condition comes from our choice of a planar undeformed entangling surface). This is a simple exercise in expanding \eqref{eq-extsurfeqn}. The result in position space is
\begin{align}
J^\mu =&-\frac{1}{2}\partial_z\gamma_{cc} \partial_z \bar{X}^\mu+\partial_a(\gamma_{ab} \partial_b \bar{X}^\mu)-\eta^{\mu\nu}\partial_z \gamma_{\nu \rho} \partial_z \bar{X}^\rho \nonumber \\
&- \eta^{\mu\nu}(\partial_a \gamma_{\nu \rho} + \partial_\rho \gamma_{\nu a} - \partial_\nu \gamma_{a \rho})\partial_a \bar{X}^\rho - \frac{1}{2}\eta^{\mu\nu}(2\partial_a \gamma_{\nu a}  - \partial_\nu \gamma_{a a})-\frac{1}{2}\partial_a \gamma_{cc} \partial_a \bar{X}^\mu.
\end{align}
$a,b,c$ indices represent the $y$-directions and repeated indices are summed over. Taking the variation and evaluating at $\bar{X}^\mu =0$ gives
\begin{align}
\delta J^\mu =&-\frac{1}{2}\partial_z \gamma_{cc} \partial_z \delta \bar{X}^\mu+\partial_a(\gamma_{ab} \partial_b \delta \bar{X}^\mu)-\eta^{\mu\nu}\partial_z \gamma_{\nu \rho} \partial_z \delta \bar{X}^\rho  \nonumber \\
&- \eta^{\mu\nu}(\partial_a \gamma_{\nu \rho} + \partial_\rho \gamma_{\nu a} - \partial_\nu \gamma_{a \rho})\partial_a \delta \bar{X}^\rho \nonumber\\
&- \frac{1}{2}\eta^{\mu\nu}(2\partial_\rho\partial_a \gamma_{\nu a}  -\partial_\rho \partial_\nu \gamma_{a a}) \delta \bar{X}^\rho-\frac{1}{2}\partial_a \gamma_{cc} \partial_a \delta \bar{X}^\mu. \label{eq-deltaJ}
\end{align}
The only terms in \eqref{eq-deltaJ} that will contribute at $k^0$ are those with two $y$ derivatives acting on $\delta \bar{X}^\mu$ or with $z$ derivatives, i.e., the first line of \eqref{eq-deltaJ}. Then the result for $\delta X_{(d)}^\mu$ at large $k$ is obtained from \eqref{eq-Xmud} as
\begin{align}
\delta X_{(d)}^\mu(k) &=  \frac{-1}{2^{d-2}\Gamma(d/2)^2}\left[\left(\ev{\gamma^{(d)\mu}_{\nu}}+\frac{1}{2}h^{ab} \ev{\gamma^{(d)}_{ab}} \eta^\mu_\nu\right)\left(\lim_{z\to 0} \frac{1}{2}z^{d} K_{d/2}(z)^2 \right)\right. \nonumber\\
&\left. ~~~~~~~~~ - \left(\eta^\mu_\nu \frac{k^ak^b}{k^2}\ev{\gamma^{(d)}_{ab}}\right)\left(\int_0^\infty dz z^{d+1}K_{d/2}(z)^2 \right)\right]e^{iky_0} \xi^\nu \nonumber\\ 
&=-\frac{8\pi G_N}{d}\left[\ev{T^\mu_\nu}+\frac{1}{2}h^{ab} \ev{T_{ab}} \eta^\mu_\nu - \frac{d}{d+1} \eta^\mu_\nu \frac{k^ak^b}{k^2}\ev{T_{ab}}\right]e^{iky_0} \xi^\nu\label{eq-Xdfinal}
\end{align}
Here we have explicitly included factors of the entangling surface metric $h^{ab}$ (which is equal to $\delta^{ab}$) rather than using repeated $a,b$ indices for added clarity. In the last line, we have used the dictionary \eqref{eq-CFTstress} to replace $\gamma^{(d)}_{\mu\nu}$ with $\ev{T_{\mu\nu}}$.

The first two terms of \eqref{eq-Xdfinal} correspond to $\delta$-functions in position space. The final term clearly contains a $\delta$-function piece which will end up being proportional to the trace of $\ev{T_{ab}}$, but it also contains off-diagonal contributions. We can use the identity
\be
\int d^{d-2}k~\frac{k^ak^b}{k^2} e^{ik(y-y_0)}  \propto \partial_a\partial_b\frac{1}{|y-y_0|^{d-4}} \propto \frac{\delta_{ab}-(d-2)(y-y_0)^a(y-y_0)^b/(y-y_0)^2}{|y-y_0|^{d-2}}.
\ee
to see the full effect in position space. However, for our purposes we are only interested in the $\delta$-function contribution. Isolating this part and combining it with the first two terms of \eqref{eq-Xdfinal}, we ultimately find
\begin{align}\label{eq-sec5upshot}
\Delta S_{\mu\nu}'' &=  2\pi \left( n_{\mu}^\rho n_\nu^\sigma \ev{T_{\rho \sigma}} + \frac{d^2-3d-2}{2(d+1)(d-2)}n_{\mu\nu} h^{ab} \ev{T_{ab}}\right)
\end{align}
where $n_{\mu\nu}$ is the normal projector of the entangling surface. This completes our derivation of \eqref{eq-conjecture}.

\section{Discussion}\label{sec:discussion}
We have found formulas for the $\delta$-function piece of the second variation of entanglement entropy in terms of the expectation values of the stress tensor. In this section we conclude by discussing a number of possible extensions and future applications of this result.


\subsection{Higher Orders in $1/N$}\label{sec-higherorder}

Since we believe \eqref{eq-nullconjecture} and \eqref{eq-conjecture} to be valid at finite $N$, it must be that our calculations are not affected by higher-order corrections within holography.

One potential source of higher-order corrections comes from incorporating quantum fluctuations in the geometry, rather than treating the geometry as a classical background. We have already addressed this issue in Section~\ref{sec:2}, but we will repeat it here. The problem of a fluctuating geometry arises because the metric fluctuation $\gamma_{\mu\nu}$ is actually a quantum operator, and as such a classical expression which is nonlinear in $\gamma_{\mu\nu}$ has an ambiguous quantum interpretation because, in general, $\ev{\gamma_{\mu\nu}^2}\neq \ev{\gamma_{\mu\nu}}^2$. However, our analysis has shown that the $\delta$-function part of the second entropy variation is determined entirely by terms which are linear in $\gamma_{\mu\nu}$, and so this problem is avoided.

There are two other classes of higher-order corrections we can consider: those coming form higher-curvature corrections to the bulk gravity, and those coming from the bulk entropy. These corrections can be encapsulated in the all-orders formula~\cite{Engelhardt:2014gca, Dong:2017aa}
\be
S = S_{\rm gen}[e(\mathcal{R})] = S_{\rm Dong}[e(\mathcal{R})]  + S_{\rm bulk}[e(R)].
\ee
The first term here is the Dong entropy functional~\cite{Dong:2013qoa}, which is an integral of geometric data over the surface $e(\mathcal{R})$,\footnote{Really $S_{\rm Dong}$ is the expectation value of geometric data, but we have already argued that it is enough to treat the geometry classically for our purposes.} and the second term is the bulk entropy lying within the region bounded by $e(\mathcal{R})$. Finally, the surface $e(\mathcal{R})$ is the one that extremizes the $S_{\rm gen}$ functional.

If we ignore the $S_{\rm bulk}$ term for a moment, then $S_{\rm Dong}$ behaves qualitatively the same way as the area in the Ryu-Takayanagi formula. The coordinates $\bar{X}^\mu$ of $e(\mathcal{R})$ obey a certain differential equation, and the variations in the entropy are still related to $\delta X_{(d)}^\mu$ as before. One change is that the overall coefficient of $\delta X_{(d)}^\mu$ relative to the entropy will change in a way that depends on the bulk higher curvature couplings. However, the dictionary relating $\gamma_{\mu\nu}$ to $T_{\mu\nu}$ also changes in a way that precisely preserves \eqref{eq-nullconjecture} and \eqref{eq-conjecture}~\cite{Akers:2017aa}.

Incorporating the $S_{\rm bulk }$ term is simple in principle but difficult in practice to deal with. Since it is $S_{\rm gen}$ that must be extremized, we have to include an extra term in the extremal surface equation of motion proportional to $\delta S_{\rm bulk}/\delta \bar{X}^\mu(y)$. That means the bulk entropy itself plays a role in determining the position of the surface. It was argued in~\cite{Akers:2016aa} (assuming some mild falloff conditions on variations of the bulk entropy) that the presence of this source could be incorporated to all orders simply by removing the explicit bulk entropy term from \eqref{eq-delta2SUV}. In other words, calculating $\delta X_{(d)}^\mu$ using the correct quantum extremal surface equation is enough to properly account for all bulk entropy contributions to the total entropy variation. At order-one in the large-$N$ expansion this prescription agrees with our analysis above, as it must. Beyond this, the most we can say about the contributions of the entropy are arguments of the type given above in Section~\ref{sec:bulkpert}. While this is a potential loophole in our arguments, we still believe that our evidence suggests that new contributions to \eqref{eq-nullconjecture} and \eqref{eq-conjecture} do not appear.


\subsection{Local Conditions On $\partial \mathcal{R}$ Are Enough}\label{sec-localconditions}

We now briefly discuss why we expect that we can relax the stationarity conditions on the entangling surface to hold just in the vicinity of the deformation point. We will focus on the null-null case, but a similar result should hold in the non-null case (where it should also be true that our restriction on expectation values for operators with $\Delta < d/2$ is allowed to be local).

We can analyze the source \eqref{eq-deltaJschematic} in a little more detail in the case where we only impose local stationarity near $y=y_0$. Even though in position space $\bU(y_0,z)$ does not contain any state-independent terms at low orders in the $z$-expansion near, the inherent non-locality of the Fourier transform $\bU(k, z)$ will contain those terms. There are two ways this could affect \eqref{eq-deltaJschematic}: through $\delta \Psi = \delta \bU$ or through the $h$-factor. In either case, the large $k$ limit reduces to the problem back to the globally-stationary setup.

For example, by setting $\delta V(k) = e^{iky_0}$ we can isolate the part of $\delta U_{(d)}$ that gives a $\delta$-function localized at $y=y_0$. Then the important part of $\delta \bar{V}$ (i.e., the state-independent part) is
\be
\delta \bar{V}(k,z) = e^{iky_0} 2^{\frac{d-2}{2}}\Gamma(d/2)(kz)^{d/2}K_{d/2}(kz).
\ee
Then we can organize \eqref{eq-deltaJschematic} as a derivative expansion of $h$, with the leading term given by
\begin{align}
\delta J(k,z) \sim e^{iky_0}h(z, y_0)(kz)^{d/2} K_{d/2}(kz),
\end{align}
and the remaining terms suppressed by powers of $k$. In other words, the integral over $k'$ in \eqref{eq-deltaJschematic} combined with the $(k-k')$-dependence of $\delta V$ essentually returns $h$ to position space localized near $y=y_0$. Only the first $d$ derivatives of $h$ at $y=y_0$ will be relevant at large $k$, so only the first $d$ derivatives of $U$ need to be set equal to zero at $y=y_0$ in order for the large-$k$ behavior to match the case where $U$ vanishes identically. Thus it is enough to have entangling surfaces which are in the $u=0$ plane up to order $d$ in $y-y_0$.

Note, this crude analysis does not strictly apply if the entangling surface cannot be globally written in terms of functions $U(y), V(y)$.  For example, an entangling surface which is topologically a sphere does not fall within the regime of our arguments. We leave an analysis of those types of regions for future work.

\subsection{Curved Backgrounds}
It is interesting to ask what happens to this proof when the boundary spacetime is curved. Our arguments make it clear that $S''_{\mu\nu}$ is completely determined by local properties of the state in the bulk and on the boundary. So naturally one would expect that there is a curved-space analogue of the same formula. In \cite{Akers:2017aa, Fu:2017ab}, several local conditions on the entangling surface and spacetime curvature were found such that the QNEC would hold in curved space and be manifestly scheme-independent. We would expect that under those same conditions one could show that $S_{vv}'' = 2\pi \ev{T_{vv}}$. Non-null variations in a curved background have yet to be explored, and it would be interesting to investigate aspects of the curved background setup in more detail.


\subsection{Connections to the QFC and Gravity}
An interesting application of our result is to the interpretation of Einstein's equations. Combining \eqref{eq-conjecture} with Einstein's equations leads to an explicit formula relating geometry to entropy. This result is the latest in a growing trend of connections between geometry and entanglement~\cite{Raamsdonk:2010aa, Maldacena:2013aa, Jacobson:2016aa, Lashkari:2013koa, Faulkner:2013aa, Swingle:2014uza, Faulkner:2017aa}.

We can make a direct connection with the deep result by Jacobson of the Einstein equation of state~\cite{Jacobson:1995aa}. There it was argued that Einstein's equations were equivalent to a statement of thermal equilibrium across an arbitrary local Rindler horizon, namely the equation $\delta Q = T\delta S$, together with an assumption that $S$ is proportional to area. This argument used a thermodynamic definition of the entropy without mentioning quantum entanglement. We can give this result a modern interpretation with the equation $S''_{vv} = 2\pi \ev{T_{vv}}$.

The connection to our result is most easily phrased in terms of the generalized entropy for a field theory coupled to gravity, which is defined as
\be
S_{\rm gen} = S_{\rm Dong} + S_{\rm ren}.
\ee
Here $G_N$ is the renormalized Newton's constant, and $S_{\rm ren}$ is the renormalized entropy of the field theory system restricted to a region, and $S_{\rm Dong}$ is the same geometric functional of the boundary of the region introduced in Section~\ref{sec-higherorder}, and which at leading order is ${\rm Area}/4G_N$. Variations of this quantity were considered in \cite{Bousso:2015mna}, where the conjecture $S''_{{\rm gen}, vv} \leq 0$ was dubbed the Quantum Focusing Conjecture (QFC).

Inspired by the arguments of \cite{Jacobson:1995aa}, we will consider evaluating $S''_{{\rm gen},vv}$ on a surface passing through a given point in an arbitrary spacetimem where $v$ now denotes a null direction of our choosing. We will want to make sure that the surface is as close to stationary as possible in the $v$ direction. It is always possible to make the expansion and shear of our surface vanish at the chosen point, but generically these quantities will have nonzero derivatives along the surface. In order to keep our calculations well-defined, and avoid potential violations of the QFC~\cite{Fu:2017aa}, we should consider deformations which are integrated over at least a Planck-sized region of the surface~\cite{Leichenauer:2017aa}. While not strictly a $\delta$-function, if the mass scales governing the matter sector are must less than the Planck scale then for all practical purposes this is the same as a $\delta$-function deformation from the point of view of the matter entropy. The result of doing this type of deformation is~\cite{Akers:2017ttv}
\be\label{eq-sgenvar}
4G_N S_{{\rm gen},vv}'' = -R_{vv} + 4G_N S_{{\rm ren},vv}+ O(\ell^2/L^4),
\ee 
where $L$ is the characteristic scale of the background geometry and $\ell$ is the Planck scale (or whatever other cutoff scale is appropriate for the effective gravitational theory). The corrections at order $\ell^2/L^4$ come both from higher curvature corrections present in $S_{\rm Dong}$ beyond the ${\rm Area}/4G_N$ term, as well as from the generic non-zero derivatives of the expansion and shear at the central point of the deformation.

Now suppose we imposed the principle that $4G_N S_{{\rm gen},vv}''$ is always of order $\ell^2/L^4$, which is much smaller than the size $1/L^2$ of the first term $-R_{vv}$. Then it must be that this large contribution is canceled by $4G_N S_{{\rm ren},vv}$, which by our result above (or, more precisely, by the appropriate curved-space generalization) is equal to $8\pi G_N \ev{T_{vv}}$. In other words, we would be imposing
\be
R_{vv} = 8\pi G_N \ev{T_{vv}} + O(\ell^2/L^4).
\ee
This is the leading-order part of the full gravitational equations of motion, up to an unknown cosmological constant term coming from our restriction to null variations. The argument can also be run the other way, so that Einstein's equations, interpreted as the leading order part of the gravitational equations of motion, become equivalent to the statement
\be
4G_N S_{{\rm gen},vv}'' = O(\ell^2/L^4).
\ee
We have essentially retraced the steps of \cite{Jacobson:1995aa}, replacing the Jacobson's original assumption of $\delta Q = T\delta S$ with the this statement about the generalized entropy, together with \eqref{eq-nullconjecture}.


\subsection{Proof for General CFTs}

We view our results as sufficient motivation to look for a proof of \eqref{eq-conjecture} and \eqref{eq-nullconjecture} in general field theories. In conformal field theories, entanglement entropy can be calculated using the replica trick. A replicated CFT is equivalent to a CFT with a twist defect. Within the technology of defect CFTs, shape deformations of entropy is generated by displacement operators (see \cite{Balakrishnan:2017aa} for a review of these concepts). The variation $\delta^2 S / \delta V(y) \delta V(y')$ then is related to the OPE structure of displacement operators in this setup. Since the coefficient of the delta function piece in \eqref{eq-variation1} is fixed to have dimension $d$ and spin 2, one might be able to see that only the stress tensor could appear as a local operator in $S_{vv}''$. It further needs to be shown that no other non-linear (in the state) contributions could appear in $S_{vv}''$. Results in that direction will be reported in future work~\cite{Arvin:2018}.

\section*{Acknowledgements}
It is a pleasure to thank Chris Akers, Raphael Bousso, Venkatesh Chandrasekaran, Thomas Faulkner, Tom Hartman, Jason Koeller, Fabio Sanches and Aron Wall for discussions. We thank Jason Koeller for collaboration in the early stages of this project.
This work is supported in part by the Berkeley Center for Theoretical Physics, by the National Science Foundation (award numbers 1214644, 1316783, and 1521446), by fqxi grant RFP3-1323, and by the US Department of Energy under Contract DE-AC02-05CH11231. The work of AL is supported by the Department of Defense (DoD) through the National Defense Science \& Engineering Graduate Fellowship (NDSEG) Program.
\appendix


\section{Connections to the ANEC}

In~\ref{sec-anecrel} we briefly review the connection between the relative entropy and the ANEC. Equation~\eqref{eq-nullconjecture} then implies an interesting connection between the off-diagonal second variation of the entropy and the ANEC. In~\ref{sec-anecpert} we analyze this result in more detail for holographic field theory states dual to perturbative bulk geometries.

\subsection{ANEC and Relative Entropy}\label{sec-anecrel}

As in Section~\ref{sec-ftsetup}, the region $\mathcal{R}$ is a region whose boundary $\partial\mathcal{R}$ lies in the $u=0$ plane. We also consider a one-parameter family of such regions, indexed by $\lambda$, with the convention that increasing $\lambda$ makes the $\mathcal{R}$ smaller. In this section we will focus on a globally pure state reduced to these regions. The relative entropy (with respect to the vacuum) and its first two derivatives obey the following set of alternating inequalities:
\be
S_{\rm rel} \geq 0, ~~~~\frac{dS_{\rm rel}}{d\lambda} \leq 0,~~~~ \frac{d^2S_{\rm rel}}{d\lambda^2} \geq 0.
\ee
The first two of these are general properties of relative entropy in quantum mechanics, known as the positivity and monotonicity of relative entropy, respectively. The third inequality is the QNEC together with strong subadditivity.

We can also consider the entropy $\bar{S}$ and relative entropy $\bar{S}_{\rm rel}$ of the complement of $\mathcal{R}$, which we will denote by $\bar{\mathcal{R}}$. Since we specified that the global state is pure, we have $\bar{S} = S$. The set of inequalities obeyed by $\bar{S}_{\rm rel}$ is 
\be
\bar{S}_{\rm rel} \geq 0, ~~~~\frac{d\bar{S}_{\rm rel}}{d\lambda} \geq 0,~~~~ \frac{d^2\bar{S}_{\rm rel}}{d\lambda^2} \geq 0.
\ee
From \eqref{eq-entderiv} and the analogous equation for $\bar{S}_{\rm rel}$, together with the monotonicity of relative entropy inequalities, we can conclude
\be\label{eq-relANEC}
\frac{d\bar{S}_{\rm rel}}{d\lambda} - \frac{dS_{\rm rel}}{d\lambda} = 2\pi \int d^{d-2}y dv~\braket{T_{vv}} \dot V(y) \geq 0.
\ee
This is the ANEC, and its connection to relative entropy was first pointed out in~\cite{Wall:2010cj, Faulkner:2016mzt}.

The relation \eqref{eq-relANEC} has interesting implications. Note that the integral of $T_{vv}$ is completely independent of $\lambda$. If we let $\lambda \to \infty$, it must be the case that $dS_{\rm rel} /d\lambda \to 0$ or else positivity of relative entropy will be violated. Similarly, as $\lambda \to -\infty$ we must have $d\bar{S}_{\rm rel} /d\lambda \to 0$. Then we can say
\be
\int_{-\infty}^\infty d\lambda~\frac{d^2S_{\rm rel}}{d\lambda^2} = \frac{dS_{\rm rel}}{d\lambda}(\infty) - \frac{dS_{\rm rel}}{d\lambda}(-\infty) = 2\pi\int d^{d-2}y dv~\braket{T_{vv}} \dot V(y).
\ee
From the definition of relative entropy, this means that
\be\label{eqn:sod}
\int_{-\infty}^{\infty} d\lambda\int d^{d-2}y~ S''\dot V(y)^2= -\int_{-\infty}^{\infty} d\lambda \int d^{d-2}yd^{d-2}y'~ \left(\frac{\delta^2 S}{\delta V(y)\delta V(y')}\right)_{\rm od}\dot V(y) \dot V(y').
\ee
So the diagonal and off-diagonal parts of the second variation entropy contribute equally when integrated over the entire one-parameter family of surface deformations. Since there are two $y$ integrals on the RHS of (\ref{eqn:sod}), na\"ively one might have thought that a limiting case for $\dot V(y)$ existed which caused the RHS of this equation to vanish while leaving the LHS finite, but this is not true. We will say more about the order-of-limits involved in the holographic context below. Applying the relation $S_{vv}'' = 2\pi \ev{T_{vv}}$ we see that, after integration, the off-diagonal variations can be related back to the ANEC:
\be\label{eq-ANECod}
2\pi\int d^{d-2}ydv~ \ev{T_{vv}} \dot V(y)= -\int_{-\infty}^{\infty} d\lambda \int d^{d-2}yd^{d-2}y'~ \left(\frac{\delta^2 S}{\delta V(y)\delta V(y')}\right)_{\rm od}\dot V(y) \dot V(y').
\ee
This is a nontrivial consequence of \eqref{eq-nullconjecture}. Note that $\delta^2 S^{\rm od}/\delta V(y)\delta V(y')\leq 0$ by strong subadditivity~\cite{Bousso:2015mna}.


\subsection{ANEC in a Perturbative Bulk}\label{sec-anecpert}

In this section we will investigate \eqref{eq-ANECod} in AdS/CFT for perturbative bulk states. Once again, we will drop the contributions of $S_{\rm bulk}$ for simplicity. This amounts to considering coherent states in the bulk.

From \eqref{eq-Sderiv}, we can see that for perturbative classical bulk states the bulk boost energy completely accounts for the off-diagonal entropy variation. Then from \eqref{eqn:Kbulkvar} we get
\be
\frac{\delta^2 S^{\rm od}}{\delta V(y_1)\delta V(y_2)} = -2\pi\left(\frac{2^{d-2}\Gamma(\frac{d-1}{2})}{\pi^{\frac{d-1}{2}}}\right)^2 \int \frac{dzd^{d-2}y}{z^{d-1}}~\ev{T_{vv}^{\rm bulk}}\frac{z^{2d}}{(z^2 + (y-y_1)^2)^{d-1}(z^2 + (y-y_2)^2)^{d-1}}
\ee
As a consequence of \eqref{eq-ANECod} we then have the equation
\be\label{eq-bulkbdryANEC}
\int d^{d-2}ydv~ \ev{T_{vv}} \dot V(y)= \int \frac{dvdzd^{d-2}y}{z^{d-1}}~\ev{T_{vv}^{\rm bulk}}\dot{\bar{V}}(y,z).
\ee
This is a nontrivial matching between the ANEC on the boundary and an associated ANEC in the bulk, made possible by the relationship between $\dot{V}$ and $\dot{\bar{V}}$ that comes from solving the extremal surface equation:
\be
\dot{\bar{V}}(y,z) = \frac{2^{d-2}\Gamma(\frac{d-1}{2})}{\pi^{\frac{d-1}{2}}}  \int d^{d-2}y' \frac{z^{d}}{(z^2 + (y-y')^2)^{d-1}}\dot{V}(y').
\ee
We can get some intuition for these equations by considering shockwave solutions in the bulk.

\paragraph{Shockwaves}
Consider a shockwave geometry in the bulk. The bulk stress tensor is~\cite{Afkhami-Jeddi:2017aa}
\be
\ev{T_{vv}^{\rm bulk}} =  Ez_0^{d-1} \delta(v) \delta^{d-2}(y)\delta(z-z_0)
\ee
and the boundary stress tensor is
\be
\ev{T_{vv}} = E\frac{2^{d-2}\Gamma\left(\frac{d-1}{2}\right)z_0^d}{\pi^{\frac{d-1}{2}}(z_0^2+y^2)^{d-1}}\delta(v)
\ee
The parameters $z_0$ and $E$ characterize the solution. One can see directly that \eqref{eq-bulkbdryANEC} holds.

It is also interesting to integrate over a finite range of the deformation parameter. As the range is extended to infinity we recover \eqref{eq-bulkbdryANEC}, but for finite amounts of deformation we can see how the diagonal and off-diagonal parts of the entropy compete. We take the undeformed surface at $\lambda=0$ to be the flat plane $V(y)=0$ and we place the shockwave at $v=v_0$. Then integrating over a range of deformations about zero we find on the boundary
\begin{align}
\int_{0}^{\lambda}d\lambda' ~\int d^{d-2}y~ \ev{T_{vv}} \dot{V}(y)^2 &= \int d^{d-2}y ~ E\frac{2^{d-2}\Gamma\left(\frac{d-1}{2}\right)z_0^d}{\pi^{\frac{d-1}{2}}(z_0^2+y^2)^{d-1}} \dot{V}(y)\Theta(\lambda \dot{V}(y=0)-v_0)\nonumber \\
&=E \dot{\bar{V}}(y=0,z=z_0)\Theta(\lambda \dot{V}(y=0)-v_0).
\end{align}
As soon as the integration range crosses $v=v_0$, the total energy jumps from zero to the final answer. On the other hand, in the bulk we get
\begin{align}
\int_{0}^{\lambda}d\lambda' ~\int \frac{dzd^{d-2}y}{z^{d-1}}~\ev{T_{vv}^{\rm bulk}}\dot{\bar{V}}(y,z)^2 &= E \dot{\bar{V}}(y=0,z=z_0)\Theta\left(\lambda \dot{\bar{V}}(y=0,z=z_0)-v_0\right).
\end{align}
This is a very similar answer, but now the jump does not occur until later: $\dot{\bar{V}}(y=0,z=z_0)$ will always be less than $\dot{V}(y)$, which means $\lambda$ has to get larger. How much larger? We can estimate it by looking at the example of a bump function deformation with $\dot{V}(y) = 1$ over a region of area $\mathcal{A}\ll z_0^{d-2}$ and zero elsewhere. Then the boundary energy will register at $\lambda=v_0$, while the bulk energy will register at
\be
\lambda =   \frac{\pi^{\frac{d-1}{2}}}{2^{d-2}\Gamma(\frac{d-1}{2})} \frac{z_0^{d-2}}{\mathcal{A}}v_0 \gg v_0~.
\ee
So for very narrow deformations, the off-diagonal contributions to the entropy can only be seen when integrated over a large range of the deformation parameter. From the boundary point of view, the parameter $z_0$ controls how diffuse the energy is in the $y$-directions. It is a measure of the nonlocality of the state. The off-diagonal entropy variations are sensitive to this nonlocality.

Note that the order of limits we have discovered here is worth repeating. If we take $\mathcal{A} \to 0$ before taking $\lambda \to \infty$ then our integration will only be sensitive to the diagonal entropy variation (i.e., the boundary stress tensor) and we will find apparent violations of \eqref{eq-ANECod}. The reason is that there are important contributions to the off-diagonal entropy variations when $\lambda \sim z_0^{d-2}/\mathcal{A}$, where $z_0$ controls the level of nonlocality in the state.

\paragraph{Superpositions of Shockwaves}

At linear order in the bulk perturbations we can take superpositions of shockwaves. This allows us to create any bulk and boundary bulk stress tensor profile along the $u=0$ plane, and in that sense represents the most general state for the purpose of this calculation. The bulk and boundary stress tensors would be
\be
\ev{T_{vv}^{\rm bulk}(y,z,v)} = z^{d-1} \rho(y,z,v)
\ee
and
\be
\braket{T_{vv}(y,v)} =  \frac{2^{d-2}\Gamma\left(\frac{d-1}{2}\right)}{\pi^{\frac{d-1}{2}}}\int d^{d-2}y' dz'  \rho(y',z',v)\frac{(z')^d}{((z')^2+(y-y')^2)^{d-1}}
\ee
The single shockwave is the special case $\rho = E\delta(v) \delta^{d-2}(y)\delta(z-z_0)$. We can repeat some of the calculations we did before, but qualitatively the results will be the same. The deformed bulk extremal surface always ``lags behind" the deformed entangling surface in a way that depends on $z$ and the width of the deformation, and as a result the bulk energy flux at finite deformation parameters will always be less than the boundary energy flux. Taking the deformation width to zero at finite deformation parameters will cause the bulk energy flux to drop to zero. It would be interesting to characterize this behavior directly in the field theory without the bulk picture.


\section{Free and Weakly-Interacting Theories}\label{sec-free}

Our conjectures \eqref{eq-conjecture} and \eqref{eq-nullconjecture} are only meant to apply to interacting theories. In this appendix we will explain how the null-null relation \eqref{eq-nullconjecture} is violated in free theories, and indicate how it might be fixed when interactions are included.

\subsection{The Case of Free Scalars}

The case of free scalar fields for entangling surfaces restricted to $u=0$ was analyzed extensively in~\cite{Bousso:2016aa}, and we will make use of that analysis here. As in Section~\ref{sec-ftsetup} we have a one-parameter family of regions indexed by $\lambda$. The deformation velocity $\dot{V}(y)$ is taken to be a unit step-function with support on a small region of area $\mathcal{A}$ in the $y$-directions. The crucial point is to focus attention on the pencil of the $u=0$ plane that is the support of $\dot{V}(y)$. As $\lambda$ varies, the entangling surface moves within this pencil but stays fixed outside of it.

\paragraph{The State and the Entropy}
For the purpose of constructing the state, we can model the full theory as a $1+1$-dimensional massless chiral boson living on the pencil, together with an auxiliary system consisting of the rest of the $u=0$ plane. This is the formalism of null quantization, which is reviewed in~\cite{Bousso:2016aa}.

There are two facts we're going to use to write down the sate $\rho(\lambda)$ on the pencil+auxiliary system. First, in the limit of small $\mathcal{A}$, the state on the pencil becomes approximately disentangled from the auxiliary system. The fully-disentangled part $\mathcal{A}^0$ part of the state looks like the vacuum, while the leading correction goes like $\mathcal{A}^{1/2}$ and consists of single-particle states on the pencil entangled with states of the auxiliary system. The second fact is that we can always translate our state in the pencil by an amount $\lambda$ so that the entangling surface is at the origin and the operators which create the state are displaced by an amount $\lambda$ from their original positions. A coordinate system where the entangling surface is fixed is preferable. Putting these facts together lets us write
\be
\rho(\lambda) = \rho_{\rm vac} \otimes \left(\sum_i e^{-2\pi K_i} |i\rangle\!\langle i | \right)+ \mathcal{A}^{1/2} \sum_{i,j} \rho_{ij}^{(1/2)}(\lambda)\otimes \left(e^{-\pi(K_i + K_j)/2} |i\rangle\!\langle j | \right)  + \cdots
\ee
The states $\ket{i}$ of the auxiliary system are merely those which diagonalize the $\mathcal{A}^0$ part of $\rho$, and the $K_i$ are numbers specifying the eigenvalues.

As indicated above the state $\rho^{(1/2)}_{ij}(\lambda)$ should be interpreted as a state on the half-line $x>0$. We can write this state in terms of a Euclidean path integral in the complex plane:
\be
\rho^{(1/2)}_{ij}[\phi^-, \phi^+] = \int_{\phi(x^+) = \phi^+}^{\phi(x^-) = \phi^-} \mathcal{D}\phi~ \mathcal{O}_{ij}(\lambda) e^{-S_{\rm E}},
\ee
where $\phi(x^\pm)$ refers to boundary conditions just above/below the positive real axis. The insertion $\mathcal{O}_{ij}(\lambda)$ is a single-field insertion which specifies the state:
\be
\mathcal{O}_{ij}(\lambda) = \int dzd\bar{z}~ \psi_{ij}(z, \bar{z}) \partial\phi(z-\lambda).
\ee
As in~\cite{Bousso:2016aa} we will normalize our field so that $\ev{\partial\phi(z)\partial\phi(0)}_{\rm vac} = -1/z^2$ and $T_{vv} = (\partial\phi)^2/4\pi \mathcal{A}$. Then one can show that $Q \equiv  S''_{vv}- 2\pi T_{vv}$ is given by
\begin{align}\label{eq-freeQ}
Q(\lambda) &= -\frac{1}{2}\sum_{ij}  \left| \int dxd\tau~(z-\lambda)^{-2+i\alpha_{ij}}   \psi_{ij}(x,\tau) \right|^2 \frac{\pi(1+\alpha_{ij}^2)\alpha_{ij}}{\sinh \pi \alpha_{ij}}e^{2\pi \alpha_{ij}}
\end{align}
where $\alpha_{ij} = K_i-K_j$ and if $z = re^{i\theta}$ with $0 \leq\theta < 2\pi$ then
\be
z^{i\alpha} = r^{i\alpha} e^{-\alpha \theta}.
\ee
The quantity $Q$ is manifestly negative, as required by the QNEC, but it is not zero.

\paragraph{Recovering the ANEC}
In Appendix~\ref{sec-anecrel} we showed how one can recover the ANEC by integrating the QNEC on a globally pure state. In the present context, we don't have any off-diagonal contributions to the entropy. Instead we have the function $Q$, and repeating the argument above would lead us to conclude
\be
\int_{-\infty}^\infty d\lambda~Q(\lambda) = -2\pi \int d\lambda ~\ev{T_{vv}(\lambda)}.
\ee
We can check this equation by integrating \eqref{eq-freeQ}. Note that the assumption of global purity that was used in Appendix~\ref{sec-anecrel} is crucial: the expectation value of $T_{vv}(\lambda)$ depends only on the part of the state proportional to $\mathcal{A}$, which we have not specified and in principle has many independent parameters. For a globally pure state there is a relationship between that part of the state and the $\mathcal{A}^{1/2}$ part of the state which we must exploit.

In the pencil+auxiliary model, the global Hilbert space consists of the full pencil plus a doubled auxiliary system. The doubling allows the auxiliary state to be purified. Let the global pure state by $\ket{\Psi}$. Then we have
\be
\ket{\Psi} = \ket{\rm vac} \otimes \left(\sum_i e^{-\pi K_i} \ket{i}\otimes\ket{i}\right)+ \mathcal{A}^{1/2}\sum_{i,j} e^{-\pi \alpha_{ij}/2} \ket{\Psi_{ij}}\otimes \ket{i}\otimes \ket{j}+\cdots
\ee
Any subsequent terms will not affect the ANEC. The factor of $\exp(-\pi \alpha_{ij}/2)$ is purely for future convenience, and the $\ket{\Psi_{ij}}$ are not necessarily normalized. The expectation value of the ANEC operator in this state is given by
\be
2\pi \int d\lambda ~\ev{T_{vv}(\lambda)}_\Psi = 2\pi \mathcal{A} \sum_{i,j} e^{-\pi \alpha_{ij}}\int d\lambda \bra{\Psi_{ij}} T_{vv}(\lambda) \ket{\Psi_{ij}}.
\ee
We can make contact with our earlier formulas by computing the density matrix $\ket{\Psi}\!\!\bra{\Psi}$ and tracing over the second copy of the auxiliary system. We find that
\be
\rho^{(1/2)}_{ij} = {\rm Tr}_{x < 0} \left(\ket{\Psi_{i j}}\!\!\bra{\rm vac}+  \ket{\rm vac}\!\!\bra{\Psi_{j i}} \right).
\ee
This lets us identify the part of $\mathcal{O}_{ij}$ in the lower half-plane as the operator which creates $\ket{\Psi_{ij}}$. Then, in our previous notation, we find
\be
2\pi \int d\lambda ~\ev{T_{vv}(\lambda)}_\Psi = 4\pi i \sum_{i,j} e^{-\pi \alpha_{ij}} \int dxd\tau dx'd\tau'   \frac{\psi_{ij}(x,\tau)\psi_{ij}(x',\tau')^*}{(z-w^*)^3}\Theta(-\tau)\Theta(-\tau').
\ee
Our job now is to reproduce this by integrating \eqref{eq-freeQ} with respect to $\lambda$. The main identity we will need is
\be
\int_{-\infty}^\infty  \frac{d\lambda}{(z-\lambda)^{2-i\alpha_{ij}}(w^*-\lambda)^{2+i\alpha_{ij}}} = \frac{4ie^{-2\pi \alpha_{ij}}\sinh\pi\alpha_{ij} }{\alpha_{ij}(1+\alpha_{ij}^2)(w^*-z)^3}\left(e^{\pi \alpha_{ij}}\Theta(\tau)\Theta(\tau') -e^{-\pi \alpha_{ij}} \Theta(-\tau)\Theta(-\tau')\right).
\ee
Using this formula, the integral of \eqref{eq-freeQ} splits into two terms. We may combine them by exchanging $i$ and $j$ in the first term, leaving us with
\begin{align}
\int d\lambda~ Q(\lambda) &= -2\pi i\sum_{ij} \int dxd\tau dx'd\tau'~ \frac{ \psi_{ij}(x,\tau)\psi_{ij}(x',\tau')^* }{(w^*-z)^3}\left(e^{\pi \alpha_{ij}}\Theta(\tau)\Theta(\tau') -e^{-\pi \alpha_{ij}} \Theta(-\tau)\Theta(-\tau')\right)  \nonumber \\
&= -4\pi i\sum_{ij} e^{-\pi \alpha_{ij}} \int dxd\tau dx'd\tau'~ \frac{ \psi_{ij}(x,\tau)\psi_{ij}(x',\tau')^* }{(z-w^*)^3}\Theta(-\tau)\Theta(-\tau').
\end{align}

\paragraph{Coherent States}
For coherent states we obtain a correspondence between $Q$ and $T_{vv}$ without integrating over $\lambda$. This must be true because coherent states satisfy $S''_{vv} = 0$, but it is reassuring to see it happen explicitly. In a coherent state of the original $d$-dimensional theory, the pencil and auxiliary system factorize and the pencil is in a $1+1$-dimensional coherent state. In other words, we have
\be
\rho(\lambda)[\phi^-, \phi^+] =\left( \int_{\phi(x^+) = \phi^+}^{\phi(x^-) = \phi^-} \mathcal{D}\phi~ e^{-S_{\rm E} + \mathcal{A}^{1/2}\mathcal{O}(\lambda)}     \right) \otimes \left(\sum_i e^{-2\pi K_i} |i\rangle\!\langle i | \right).
\ee
We can obtain $Q$ for this state by taking the general equation \eqref{eq-freeQ} specializing to the case where $\psi_{ij} = \psi \delta_{ij}\exp(-\pi K_i)$. Making use of the normalization condition $\sum_i \exp(-2\pi K_i) = 1$ we find the simple expression
\begin{align}\label{eq-coherentQ}
Q_{\rm coherent}(\lambda) &= -\frac{1}{2}\left| \int dxd\tau~\frac{ \psi(x,\tau)}{(z-\lambda)^2} \right|^2 = - \frac{1}{2\mathcal{A}}\ev{\partial\phi(\lambda)}_{\rm coherent}^2.
\end{align}
We recognize this as simply $-2\pi \ev{T_{vv}}_{\rm coherent}$, as expected.


\subsection{Weakly Interacting Effective Field Theories}\label{sec-weak}

In the main text we provided evidence for that $S''_{vv} = 2\pi \ev{T_{vv}}$ for interacting theories, but in the previous section we explained that for free theories $Q = S''_{vv} - 2\pi \ev{T_{vv}}$ was nonzero, and in fact could be quite large. In this section we will show how we can transition from $S''_{vv} \neq 2\pi \ev{T_{vv}}$ to $S''_{vv} = 2\pi \ev{T_{vv}}$ when a weak coupling is turned on.\footnote{We thank Thomas Faulkner for first pointing out the arguments we present in this section.}

The essential point is that one should always consider the total variation $d^2S/d\lambda^2$ as the primary physical quantity. $S''_{vv}$ is a derived quantity obtained by considering a limiting case of arbitrarily thin deformations. However, a weakly-coupled effective field theory in the IR comes with a cutoff scale $\epsilon$, and we cannot reliably compute $d^2 S /d\lambda^2$ for deformations of width $\ell \lesssim \epsilon$. Now we will see how this can resolve the issue.

In the free theory, as we have explained above, the second functional derivative of the entropy has the form
\be
\frac{\delta^2 S_{\rm free}}{\delta V(y)\delta V(y')} = 2\pi \ev{T_{vv}}\delta^{(d-2)}(y-y') + Q\delta^{(d-2)}(y-y') + \left(\frac{\delta^2 S}{\delta V(y)\delta V(y')}\right)_{\rm od}.
\ee
The function $Q$ is related to the square of the expectation value of the field $\partial \phi$. This is especially obvious in the formula for the coherent state, \eqref{eq-coherentQ}, but the more general formula is essentially of the same form. In a free theory $(\partial \phi)^2$ has dimension $d$ and is exactly of the right form to contribute to a $\delta$-function. This fact was touched upon in the Introduction. When we turn on a weak coupling $g$, the dimension of $\phi$ will shift to $\Delta_\phi = (d-2)/2 + \gamma(g)$.\footnote{We treat $g$ and $\gamma$ as fixed numbers that do not themselves depend on scale. A more complete treatment that incorporates the RG flow of the coupling would be interesting.} There will still be a term in the second variation of the entropy associated to $(\partial \phi)^2$, which we will call $Q_g$, but now it no longer comes with a $\delta$-function:
\be\label{eq-gvar}
\frac{\delta^2 S_g}{\delta V(y)\delta V(y')} = 2\pi \braket{T_{vv}}\delta^{(d-2)}(y-y') + Q_gf_g(y-y') + \left(\small{\text{other off-diagonal terms}}\right).
\ee
Here $f_g$ is some function of mass dimension $d-2-2\gamma$ which limits to a $\delta$-function as $g\to 0$, such as $f_g(y)\sim \gamma/ y^{d-2-2\gamma}$. So the $Q_g$ term has migrated from the $\delta$-function to the off-diagonal part of the entropy variation. 

Now consider integrating \eqref{eq-gvar} twice against a deformation profile of width $\ell$ and unit height to get a total second derivative of the entropy. Suppose that $\ell$ is very small compared to the length scales of the state, but still large compared to the cutoff $\epsilon$. Then we have 
\be
\frac{d^2S_g}{d\lambda^2} = 2\pi \braket{T_{vv}}\ell^{d-2} + Q_g \ell^{d-2+2\gamma} + \left(\small{\text{other smeared off-diagonal terms}}\right).
\ee 
We can write $Q_g\sim Q M^{2\gamma}$, where $M$ is a mass scale characterizing the state and $Q$ is what we get in the $g\to0$ limit. So at weak coupling, we can say that
\be
Q_g \ell^{d-2+2\gamma} \sim Q\ell^{d-2}\left(1 + 2\gamma \log M\ell +\cdots\right).
\ee
Thus we find that the answer for the weakly-coupled theory is approximately the same as for the free theory, as long as $\gamma \log M\ell \ll 1$. The smallest we can make $\ell$ is of order the cutoff $\epsilon$, and the condition that $\gamma \log M\epsilon$ remain small is analogous to the problem of large logarithms in perturbation theory. The renormalization group is typically used to get around the problem of large logarithms, and it would be interesting to apply those same ideas to the present situation.

This argument hints that for general effective field theories $S''_{vv}$ may not have a good operational meaning in terms of physical observables. The relevant condition for isolating the $\delta$-function is that $(M\ell)^{2\gamma} \ll 1$ should be possible within the effective description. Clearly this can be done in an exact CFT with finite anomalous dimensions, but it should also be possible if the theory is approximately given by an interacting CFT over some large range of length scales. For instance, if an interacting CFT is weakly coupled to gravity and we consider states with energy $M$ much less than the Planck scale then it should be possible to have $(M\ell)^{2\gamma} \ll 1$ while maintaining $\ell \gg \ell_{\rm Planck}$.

Finally, a more precise version of the arguments given above can be given by interpreting the second functional derivative of the entropy as an OPE. We hope to use these techniques to find the exact form of $f_g$ in future work \cite{Arvin:2018}.

\bibliographystyle{utcaps}
\bibliography{all}

\providecommand{\href}[2]{#2}\begingroup\raggedright\begin{thebibliography}{10}

\bibitem{Casini:2008aa}
H.~Casini, ``Relative entropy and the Bekenstein bound,'' {\em
  Class.Quant.Grav.} {\bfseries 25} (2008) 205021,
  \href{http://arxiv.org/abs/0804.2182}{{\ttfamily 0804.2182}}.
  \url{https://arxiv.org/abs/0804.2182}.

\bibitem{Bousso:2014sda}
R.~Bousso, H.~Casini, Z.~Fisher, and J.~Maldacena, ``{Proof of a Quantum Bousso
  Bound},'' \href{http://dx.doi.org/10.1103/PhysRevD.90.044002}{{\em Phys.
  Rev.} {\bfseries D90} no.~4, (2014) 044002},
\href{http://arxiv.org/abs/1404.5635}{{\ttfamily arXiv:1404.5635 [hep-th]}}.

\bibitem{Bousso:2015aa}
R.~Bousso, H.~Casini, Z.~Fisher, and J.~Maldacena, ``Entropy on a null surface
  for interacting quantum field theories and the Bousso bound,'' {\em Phys.
  Rev. D} {\bfseries 91} (2015) 084030,
  \href{http://arxiv.org/abs/1406.4545}{{\ttfamily 1406.4545}}.
  \url{https://arxiv.org/abs/1406.4545}.

\bibitem{Faulkner:2016mzt}
T.~Faulkner, R.~G. Leigh, O.~Parrikar, and H.~Wang, ``{Modular Hamiltonians for
  Deformed Half-Spaces and the Averaged Null Energy Condition},''
\href{http://arxiv.org/abs/1605.08072}{{\ttfamily arXiv:1605.08072 [hep-th]}}.

\bibitem{Hartman:2016lgu}
T.~Hartman, S.~Kundu, and A.~Tajdini, ``{Averaged Null Energy Condition from
  Causality},''
\href{http://arxiv.org/abs/1610.05308}{{\ttfamily arXiv:1610.05308 [hep-th]}}.

\bibitem{Bousso:2016aa}
R.~Bousso, Z.~Fisher, J.~Koeller, S.~Leichenauer, and A.~C. Wall, ``Proof of
  the Quantum Null Energy Condition,'' {\em Phys. Rev. D} {\bfseries 93} (2016)
  024017, \href{http://arxiv.org/abs/1509.02542}{{\ttfamily 1509.02542}}.
  \url{https://arxiv.org/abs/1509.02542}.

\bibitem{Koeller:2015qmn}
J.~Koeller and S.~Leichenauer, ``{Holographic Proof of the Quantum Null Energy
  Condition},''
\href{http://arxiv.org/abs/1512.06109}{{\ttfamily arXiv:1512.06109 [hep-th]}}.

\bibitem{Balakrishnan:2017aa}
S.~Balakrishnan, T.~Faulkner, Z.~U. Khandker, and H.~Wang, ``A General Proof of
  the Quantum Null Energy Condition,''
  \href{http://arxiv.org/abs/1706.09432}{{\ttfamily 1706.09432}}.
  \url{https://arxiv.org/abs/1706.09432}.

\bibitem{Wall:2017aa}
A.~C. Wall, ``A Lower Bound on the Energy Density in Classical and Quantum
  Field Theories,'' {\em Phys. Rev. Lett.} {\bfseries 118} (2017) 151601,
  \href{http://arxiv.org/abs/1701.03196}{{\ttfamily 1701.03196}}.
  \url{https://arxiv.org/abs/1701.03196}.

\bibitem{Ecker:2017aa}
C.~Ecker, D.~Grumiller, W.~van~der Schee, and P.~Stanzer, ``Saturation of the
  Quantum Null Energy Condition in Far-From-Equilibrium Systems,''
  \href{http://arxiv.org/abs/1710.09837}{{\ttfamily 1710.09837}}.
  \url{https://arxiv.org/abs/1710.09837}.

\bibitem{Casini:2017aa}
H.~Casini, E.~Teste, and G.~Torroba, ``Modular Hamiltonians on the null plane
  and the Markov property of the vacuum state,''
  \href{http://arxiv.org/abs/1703.10656}{{\ttfamily 1703.10656}}.
  \url{https://arxiv.org/abs/1703.10656}.

\bibitem{Akers:2017aa}
C.~Akers, V.~Chandrasekaran, S.~Leichenauer, A.~Levine, and
  A.~Shahbazi-Moghaddam, ``The Quantum Null Energy Condition, Entanglement
  Wedge Nesting, and Quantum Focusing,''
  \href{http://arxiv.org/abs/1706.04183}{{\ttfamily 1706.04183}}.
  \url{https://arxiv.org/abs/1706.04183}.

\bibitem{Fu:2017aa}
Z.~Fu, J.~Koeller, and D.~Marolf, ``Violating the Quantum Focusing Conjecture
  and Quantum Covariant Entropy Bound in $d\ge 5$ dimensions,'' {\em Class.
  Quantum Grav.} {\bfseries 34} (2017) 175006,
  \href{http://arxiv.org/abs/1705.03161}{{\ttfamily 1705.03161}}.
  \url{https://arxiv.org/abs/1705.03161}.

\bibitem{Jacobson:1995aa}
T.~Jacobson, ``Thermodynamics of Spacetime: The Einstein Equation of State,''
  {\em Phys.Rev.Lett.} {\bfseries 75} (1995) 1260--1263,
  \href{http://arxiv.org/abs/gr-qc/9504004}{{\ttfamily gr-qc/9504004}}.
  \url{https://arxiv.org/abs/gr-qc/9504004}.

\bibitem{Koeller:2016aa}
J.~Koeller and S.~Leichenauer, ``Holographic Proof of the Quantum Null Energy
  Condition,'' {\em Phys. Rev. D} {\bfseries 94} (2016) 024026,
  \href{http://arxiv.org/abs/1512.06109}{{\ttfamily 1512.06109}}.
  \url{https://arxiv.org/abs/1512.06109}.

\bibitem{Bousso:2015mna}
R.~Bousso, Z.~Fisher, S.~Leichenauer, and A.~C. Wall, ``{A Quantum Focussing
  Conjecture},''
\href{http://arxiv.org/abs/1506.02669}{{\ttfamily arXiv:1506.02669 [hep-th]}}.

\bibitem{Hung:2011ta}
L.-Y. Hung, R.~C. Myers, and M.~Smolkin, ``{Some Calculable Contributions to
  Holographic Entanglement Entropy},''
  \href{http://dx.doi.org/10.1007/JHEP08(2011)039}{{\em JHEP} {\bfseries 08}
  (2011) 039},
\href{http://arxiv.org/abs/1105.6055}{{\ttfamily arXiv:1105.6055 [hep-th]}}.

\bibitem{deHaro:2000vlm}
S.~de~Haro, S.~N. Solodukhin, and K.~Skenderis, ``{Holographic reconstruction
  of space-time and renormalization in the AdS / CFT correspondence},''
  \href{http://dx.doi.org/10.1007/s002200100381}{{\em Commun. Math. Phys.}
  {\bfseries 217} (2001) 595--622},
\href{http://arxiv.org/abs/hep-th/0002230}{{\ttfamily arXiv:hep-th/0002230
  [hep-th]}}.

\bibitem{Marolf:2016aa}
D.~Marolf and A.~C. Wall, ``State-Dependent Divergences in the Entanglement
  Entropy,'' {\em JHEP} {\bfseries 10} (2016) 109,
  \href{http://arxiv.org/abs/1607.01246}{{\ttfamily 1607.01246}}.
  \url{https://arxiv.org/abs/1607.01246}.

\bibitem{Ryu:2006bv}
S.~Ryu and T.~Takayanagi, ``{Holographic derivation of entanglement entropy
  from AdS/CFT},'' \href{http://dx.doi.org/10.1103/PhysRevLett.96.181602}{{\em
  Phys. Rev. Lett.} {\bfseries 96} (2006) 181602},
\href{http://arxiv.org/abs/hep-th/0603001}{{\ttfamily arXiv:hep-th/0603001
  [hep-th]}}.

\bibitem{Hubeny:2007xt}
V.~E. Hubeny, M.~Rangamani, and T.~Takayanagi, ``{A Covariant holographic
  entanglement entropy proposal},''
  \href{http://dx.doi.org/10.1088/1126-6708/2007/07/062}{{\em JHEP} {\bfseries
  07} (2007) 062},
\href{http://arxiv.org/abs/0705.0016}{{\ttfamily arXiv:0705.0016 [hep-th]}}.

\bibitem{Faulkner:2013ana}
T.~Faulkner, A.~Lewkowycz, and J.~Maldacena, ``{Quantum Corrections to
  Holographic Entanglement Entropy},''
  \href{http://dx.doi.org/10.1007/JHEP11(2013)074}{{\em JHEP} {\bfseries 11}
  (2013) 074},
\href{http://arxiv.org/abs/1307.2892}{{\ttfamily arXiv:1307.2892 [hep-th]}}.

\bibitem{Engelhardt:2014aa}
N.~Engelhardt and A.~C. Wall, ``Quantum Extremal Surfaces: Holographic
  Entanglement Entropy beyond the Classical Regime,''
  \href{http://arxiv.org/abs/1408.3203}{{\ttfamily 1408.3203}}.
  \url{https://arxiv.org/abs/1408.3203}.

\bibitem{Dong:2017aa}
X.~Dong and A.~Lewkowycz, ``Entropy, Extremality, Euclidean Variations, and the
  Equations of Motion,'' \href{http://arxiv.org/abs/1705.08453}{{\ttfamily
  1705.08453}}. \url{https://arxiv.org/abs/1705.08453}.

\bibitem{Klebanov:1999tb}
I.~R. Klebanov and E.~Witten, ``{AdS / CFT correspondence and symmetry
  breaking},'' \href{http://dx.doi.org/10.1016/S0550-3213(99)00387-9}{{\em
  Nucl. Phys.} {\bfseries B556} (1999) 89--114},
\href{http://arxiv.org/abs/hep-th/9905104}{{\ttfamily arXiv:hep-th/9905104
  [hep-th]}}.

\bibitem{Balasubramanian:aa}
V.~Balasubramanian and P.~Kraus, ``A Stress Tensor for Anti-de Sitter
  Gravity,'' \href{http://arxiv.org/abs/hep-th/9902121}{{\ttfamily
  hep-th/9902121}}. \url{https://arxiv.org/abs/hep-th/9902121}.

\bibitem{Jafferis:2015del}
D.~L. Jafferis, A.~Lewkowycz, J.~Maldacena, and S.~J. Suh, ``{Relative entropy
  equals bulk relative entropy},''
  \href{http://dx.doi.org/10.1007/JHEP06(2016)004}{{\em JHEP} {\bfseries 06}
  (2016) 004},
\href{http://arxiv.org/abs/1512.06431}{{\ttfamily arXiv:1512.06431 [hep-th]}}.

\bibitem{Dong:2013qoa}
X.~Dong, ``{Holographic Entanglement Entropy for General Higher Derivative
  Gravity},'' \href{http://dx.doi.org/10.1007/JHEP01(2014)044}{{\em JHEP}
  {\bfseries 01} (2014) 044},
\href{http://arxiv.org/abs/1310.5713}{{\ttfamily arXiv:1310.5713 [hep-th]}}.

\bibitem{Engelhardt:2014gca}
N.~Engelhardt and A.~C. Wall, ``{Quantum Extremal Surfaces: Holographic
  Entanglement Entropy Beyond the Classical Regime},''
  \href{http://dx.doi.org/10.1007/JHEP01(2015)073}{{\em JHEP} {\bfseries 01}
  (2015) 073},
\href{http://arxiv.org/abs/1408.3203}{{\ttfamily arXiv:1408.3203 [hep-th]}}.

\bibitem{Nozaki:2013aa}
M.~Nozaki, T.~Numasawa, A.~Prudenziati, and T.~Takayanagi, ``Dynamics of
  Entanglement Entropy from Einstein Equation,''
  \href{http://arxiv.org/abs/1304.7100}{{\ttfamily 1304.7100}}.
  \url{https://arxiv.org/abs/1304.7100}.

\bibitem{Wall:2011hj}
A.~C. Wall, ``{A proof of the generalized second law for rapidly changing
  fields and arbitrary horizon slices},''
  \href{http://dx.doi.org/10.1103/PhysRevD.87.069904,
  10.1103/PhysRevD.85.104049}{{\em Phys. Rev.} {\bfseries D85} (2012) 104049},
  \href{http://arxiv.org/abs/1105.3445}{{\ttfamily arXiv:1105.3445 [gr-qc]}}.
[Erratum: Phys. Rev.D87,no.6,069904(2013)].

\bibitem{Faulkner:2016aa}
T.~Faulkner, R.~G. Leigh, and O.~Parrikar, ``Shape Dependence of Entanglement
  Entropy in Conformal Field Theories,'' {\em JHEP} {\bfseries 1604} (2016)
  088, \href{http://arxiv.org/abs/1511.05179}{{\ttfamily 1511.05179}}.
  \url{https://arxiv.org/abs/1511.05179}.

\bibitem{Bhattacharya:2014vja}
J.~Bhattacharya, V.~E. Hubeny, M.~Rangamani, and T.~Takayanagi, ``{Entanglement
  Density and Gravitational Thermodynamics},''
  \href{http://dx.doi.org/10.1103/PhysRevD.91.106009}{{\em Phys. Rev.}
  {\bfseries D91} no.~10, (2015) 106009},
\href{http://arxiv.org/abs/1412.5472}{{\ttfamily arXiv:1412.5472 [hep-th]}}.

\bibitem{Akers:2016aa}
C.~Akers, J.~Koeller, S.~Leichenauer, and A.~Levine, ``Geometric Constraints
  from Subregion Duality Beyond the Classical Regime,''
  \href{http://arxiv.org/abs/1610.08968}{{\ttfamily 1610.08968}}.
  \url{https://arxiv.org/abs/1610.08968}.

\bibitem{Fu:2017ab}
Z.~Fu and D.~Marolf, ``A bare quantum null energy condition,''
  \href{http://arxiv.org/abs/1711.02330}{{\ttfamily 1711.02330}}.
  \url{https://arxiv.org/abs/1711.02330}.

\bibitem{Raamsdonk:2010aa}
M.~V. Raamsdonk, ``Building up spacetime with quantum entanglement,'' {\em
  Gen.Rel.Grav.42:2323-2329,2010; Int.J.Mod.Phys.D} {\bfseries 19} (2010)
  2429--2435, \href{http://arxiv.org/abs/1005.3035}{{\ttfamily 1005.3035}}.
  \url{https://arxiv.org/abs/1005.3035}.

\bibitem{Maldacena:2013aa}
J.~Maldacena and L.~Susskind, ``Cool horizons for entangled black holes,''
  \href{http://arxiv.org/abs/1306.0533}{{\ttfamily 1306.0533}}.
  \url{https://arxiv.org/abs/1306.0533}.

\bibitem{Jacobson:2016aa}
T.~Jacobson, ``Entanglement Equilibrium and the Einstein Equation,'' {\em Phys.
  Rev. Lett.} {\bfseries 116} (2016) 201101,
  \href{http://arxiv.org/abs/1505.04753}{{\ttfamily 1505.04753}}.
  \url{https://arxiv.org/abs/1505.04753}.

\bibitem{Lashkari:2013koa}
N.~Lashkari, M.~B. McDermott, and M.~Van~Raamsdonk, ``{Gravitational Dynamics
  from Entanglement `Thermodynamics'},''
  \href{http://dx.doi.org/10.1007/JHEP04(2014)195}{{\em JHEP} {\bfseries 04}
  (2014) 195},
\href{http://arxiv.org/abs/1308.3716}{{\ttfamily arXiv:1308.3716 [hep-th]}}.

\bibitem{Faulkner:2013aa}
T.~Faulkner, M.~Guica, T.~Hartman, R.~C. Myers, and M.~V. Raamsdonk,
  ``Gravitation from Entanglement in Holographic CFTs,''
  \href{http://arxiv.org/abs/1312.7856}{{\ttfamily 1312.7856}}.
  \url{https://arxiv.org/abs/1312.7856}.

\bibitem{Swingle:2014uza}
B.~Swingle and M.~Van~Raamsdonk, ``{Universality of Gravity from
  Entanglement},''
\href{http://arxiv.org/abs/1405.2933}{{\ttfamily arXiv:1405.2933 [hep-th]}}.

\bibitem{Faulkner:2017aa}
T.~Faulkner, F.~M. Haehl, E.~Hijano, O.~Parrikar, C.~Rabideau, and M.~V.
  Raamsdonk, ``Nonlinear Gravity from Entanglement in Conformal Field
  Theories,'' \href{http://arxiv.org/abs/1705.03026}{{\ttfamily 1705.03026}}.
  \url{https://arxiv.org/abs/1705.03026}.

\bibitem{Leichenauer:2017aa}
S.~Leichenauer, ``The Quantum Focusing Conjecture Has Not Been Violated,''
  \href{http://arxiv.org/abs/1705.05469}{{\ttfamily 1705.05469}}.
  \url{https://arxiv.org/abs/1705.05469}.

\bibitem{Akers:2017ttv}
C.~Akers, V.~Chandrasekaran, S.~Leichenauer, A.~Levine, and
  A.~Shahbazi-Moghaddam, ``{The Quantum Null Energy Condition, Entanglement
  Wedge Nesting, and Quantum Focusing},''
\href{http://arxiv.org/abs/1706.04183}{{\ttfamily arXiv:1706.04183 [hep-th]}}.

\bibitem{Arvin:2018}
S.~Balakrishnan, V.~Chandreskaran, T.~Faulkner, A.~Levine, and
  A.~Shahbazi-Moghaddam {\em To Appear} (2018) .

\bibitem{Wall:2010cj}
A.~C. Wall, ``{A Proof of the Generalized Second Law for Rapidly-Evolving
  Rindler Horizons},'' \href{http://dx.doi.org/10.1103/PhysRevD.82.124019}{{\em
  Phys. Rev.} {\bfseries D82} (2010) 124019},
\href{http://arxiv.org/abs/1007.1493}{{\ttfamily arXiv:1007.1493 [gr-qc]}}.

\bibitem{Afkhami-Jeddi:2017aa}
N.~Afkhami-Jeddi, T.~Hartman, S.~Kundu, and A.~Tajdini, ``Shockwaves from the
  Operator Product Expansion,''
  \href{http://arxiv.org/abs/1709.03597}{{\ttfamily 1709.03597}}.
  \url{https://arxiv.org/abs/1709.03597}.

\end{thebibliography}\endgroup

\end{document}